\documentclass{emulateapj}

\usepackage[usenames]{color}
\definecolor{darkred}{rgb}{0.75,0.0,0.01}

\def\arcs{$''$}
\def\mum{$\mu m$}

\def\BB{$B_{435}$}
\def\vv{$v_{606}$}
\def\ii{$i_{775}$}
\def\zz{$z_{850}$}
\def\JJ{$J_{110}$}
\def\HH{$H_{160}$}

\def\zdrops{$z_{850}-$dropouts}

\newcommand{\udfone}{UDF-640-1417}
\newcommand{\udftwo}{UDF-983-964}
\newcommand{\udfthree}{UDF-387-1125}
\newcommand{\udffour}{UDF-3244-4727}
\newcommand{\gnsone}{GNS-zD1}
\newcommand{\gnstwo}{GNS-zD2}
\newcommand{\gnsthree}{GNS-zD3}
\newcommand{\gnsfour}{GNS-zD4}
\newcommand{\gnsfive}{GNS-zD5}
\newcommand{\cdfone}{CDFS-3225-4627}
\newcommand{\hdfone}{HDFN-3654-1216}

\def\figstamps{
\begin{figure*}[h]
\centering
\includegraphics[width=0.60\textwidth]{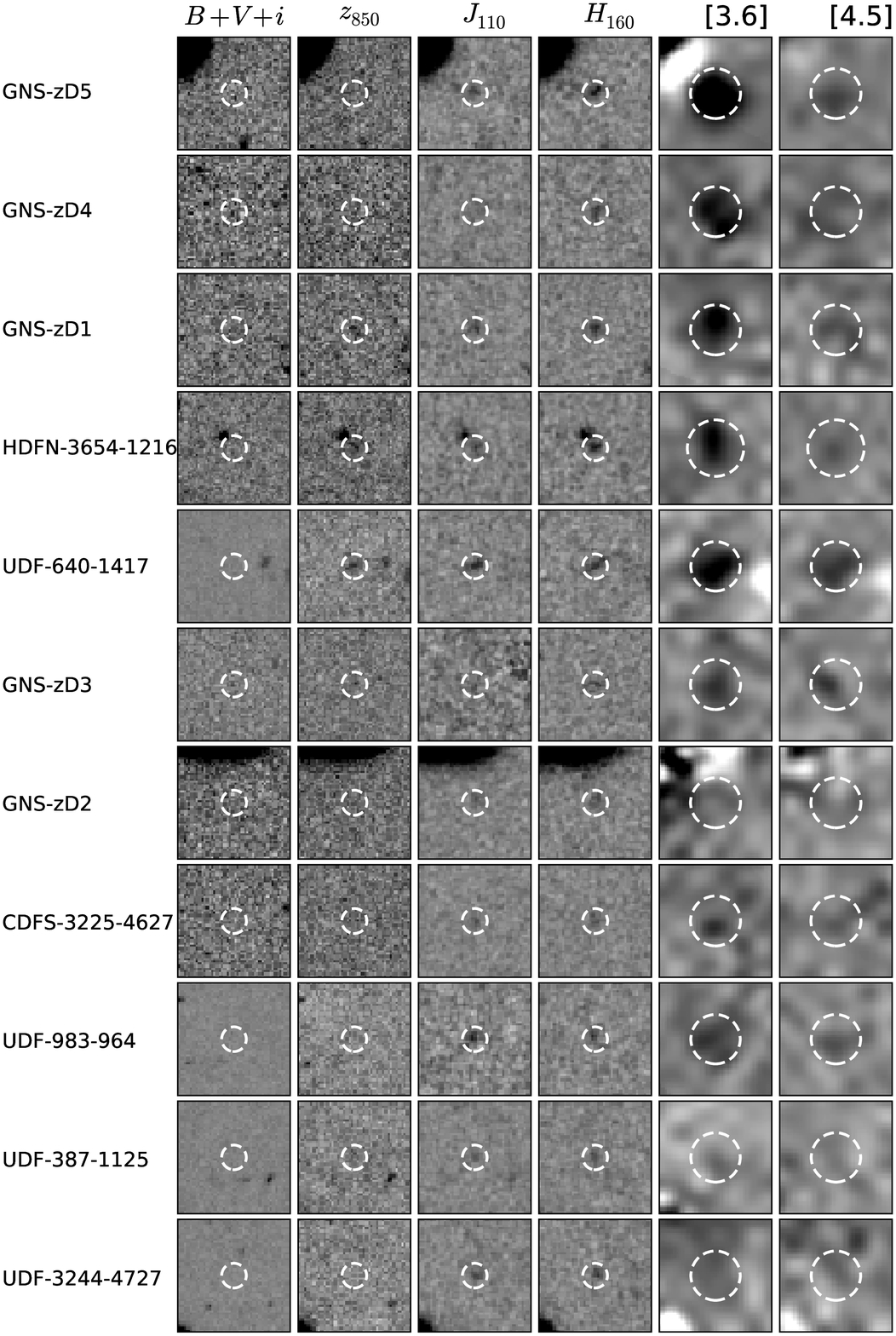}

\caption{Postage stamps of the \zdrops. Each stamp is 4\arcs$\times$4\arcs~in
size ($\sim21$ kpc at $z=7$). The dashed circles indicate the apertures used for
photometry (0.9\arcs\ for the optical to NIR and 1.8\arcs\ in the IRAC bands).
All the sources are undetected in \BB, \vv, and \ii~and only \udfone\ and
\hdfone\ are weakly detected in \zz. The latest two sources are estimated to be
the lowest redshifts of the sample. 3.6 \mum\ and 4.5 \mum\ stamps show the
sources after the flux from the nearby neighbors have been fit and removed
(i.e., they are ``cleaned'' images), with the two epochs of IRAC data added
together when available. The cleaning process in the case of
sources \gnsfive\ and \gnstwo\ has left nearby residuals (seen in white)
attributed to the close and bright nearby sources visible in the other bands. To
a lesser degree this seems to also be the case for sources \udfone\ and \udffour.
Since these latter sources were observed in both epochs, it was possible to
check that the measured fluxes are consistent among them. As expected, all
sources are undetected in the 5.8 and 8.0 \mum\ channels. The sources have been
placed in order of increasing \HH\ magnitude from top to bottom.}
\label{stamps} 
\end{figure*}
}

\def\fighist{
\begin{figure}[h]
\centering
\includegraphics[width=0.5 \textwidth]{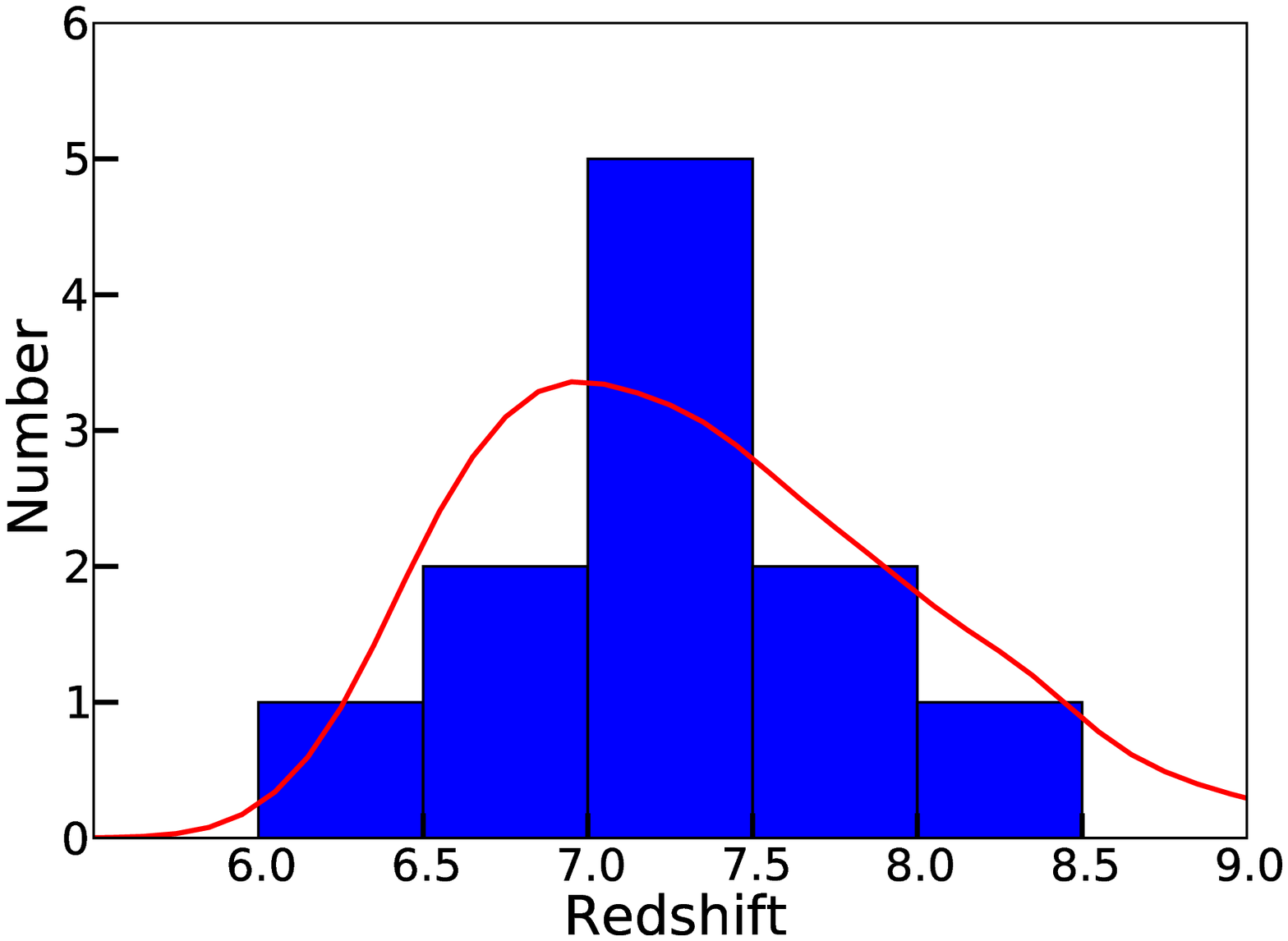}

\caption{The photometric redshifts of our $z\sim7$ $z$-dropout sample 
(histogram).  The solid line shows the redshift distribution that 
\cite{Bouwens2008} predict for the $z$-dropout selection.  The photometric 
redshifts were estimated with EAZY (see \S4).  Typical uncertainties in the 
redshift for individual sources is $\Delta z \sim0.6$.  The lowest redshift in 
the sample corresponds to object \hdfone\ at $z=6.2$.  This object presents the 
bluest \zz $-$ \JJ\ color due to it being weakly detected in the \zz\ band 
image.}
\label{zdist} 
\end{figure}
}

\def\figseds{
\begin{figure*}[p]
\centering
\includegraphics[totalheight=0.9\textheight]{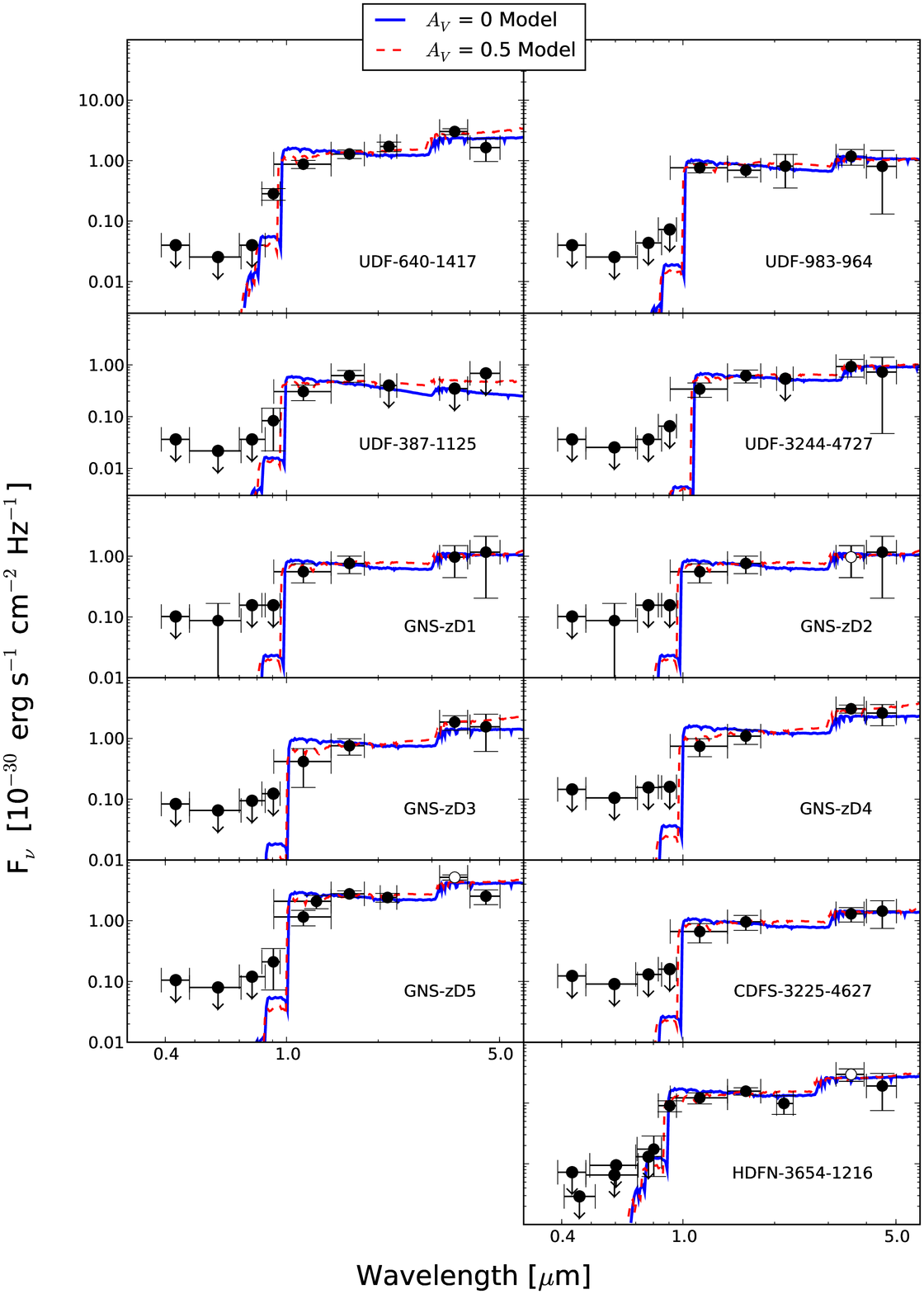}

\caption{Observed SEDs and synthetic stellar population model fits. The error
bars and upper limits are 1$\sigma$. The models shown here are from
\citet{Bruzual2003} with a Salpeter IMF and 0.2 Z$_\odot$. The red curve
represents the best fit models without extinction and the blue curve shows the
effect of imposing a maximal extinction of $A_V=0.5$. The open circles shown for
the 3.6 \mum\ flux measurements of objects \gnstwo, \gnsfive, and \hdfone\
correspond to points with poor photometry.}
\label{seds1}
\end{figure*}
}

\def\figmeansed{
\begin{figure}[ht]
\centering
\includegraphics[width=0.47\textwidth]{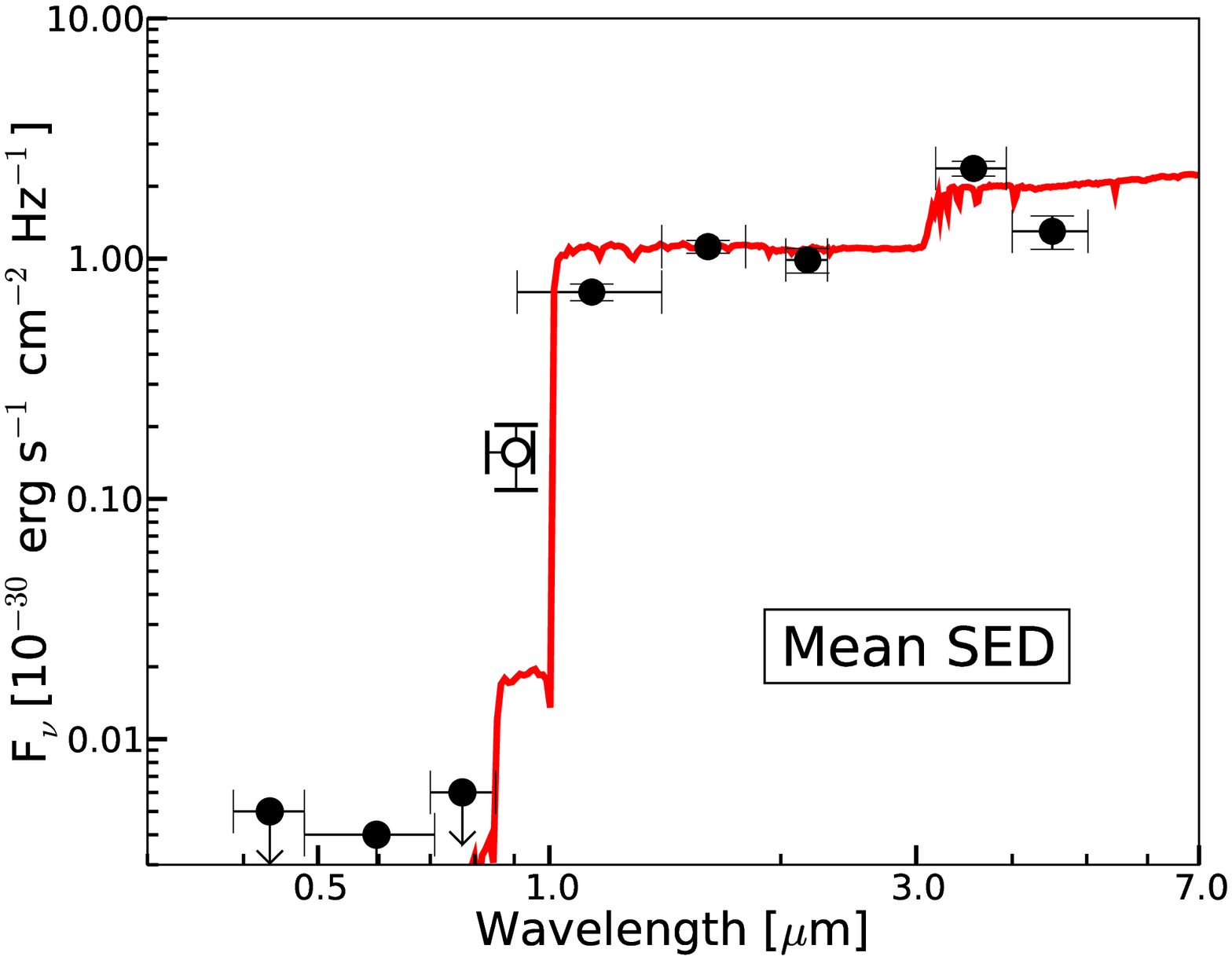}

\caption{The average observed SED (black circles) for galaxies derived from the
11 sources in our $z$-dropout selection. Error bars and upper limits are at
$1\sigma$. The fluxes of each source in our $z\sim7$ $z$-dropout selection was
scaled so that their \HH-band flux matched the average \HH-band flux for the
sample, after which the rescaled fluxes from the entire sample were averaged.
Flux measurements which were poor (due to difficulties in precisely subtracting
a nearby neighbor, as for the 3.6\mum\ flux measurements of \gnstwo, \gnsfive,
and \hdfone) were not included in the average. Note, the average $K$-band flux
we derive here was only determined from the 6 sources for which we had deep
$K$-band data. Stellar population models were fit to this average observed SED
using the same technique that what was used for individual sources (model shown
in solid line). The parameters of the fit are presented in table \ref{fits}. The
average \zz-band flux (open circle) was not included in the fit -- since it
would not make sense to include both the \zz\ and \JJ-band flux (both of which
have a different dependence on redshift) in the fits. This average SED shows no
detection in the optical $B$, $V$, or $i$ bands and a very large break ($>$3
mag) between the optical and $J$ bands -- suggesting that the majority of
sources in our sample do in fact correspond to $z\sim7$ galaxies. The extremely
blue $H-K$ color, which traces the UV continuum, imply that the dust reddening
must be very low -- consistent with our assumptions in modeling individual
$z$-dropouts (\S6). The pronounced break ($\sim$1 mag) between the K and the 3.6
\mum\ bands suggests the presence of a Balmer break, indicating that the typical
$z\sim7$ galaxy has experienced several previous generations of star formation.
We have used the derived M/L ratio from this mean SED to make a simplified
estimate of the stellar mass density of the universe at $z\sim7$ of 4.5$\times
10^{5}M_\odot Mpc^{-3}$ (see \S7).}
\label{meansed} 
\end{figure}
}

\def\figSSFRz{
\begin{figure}[ht]
\centering
\includegraphics[width=0.47\textwidth]{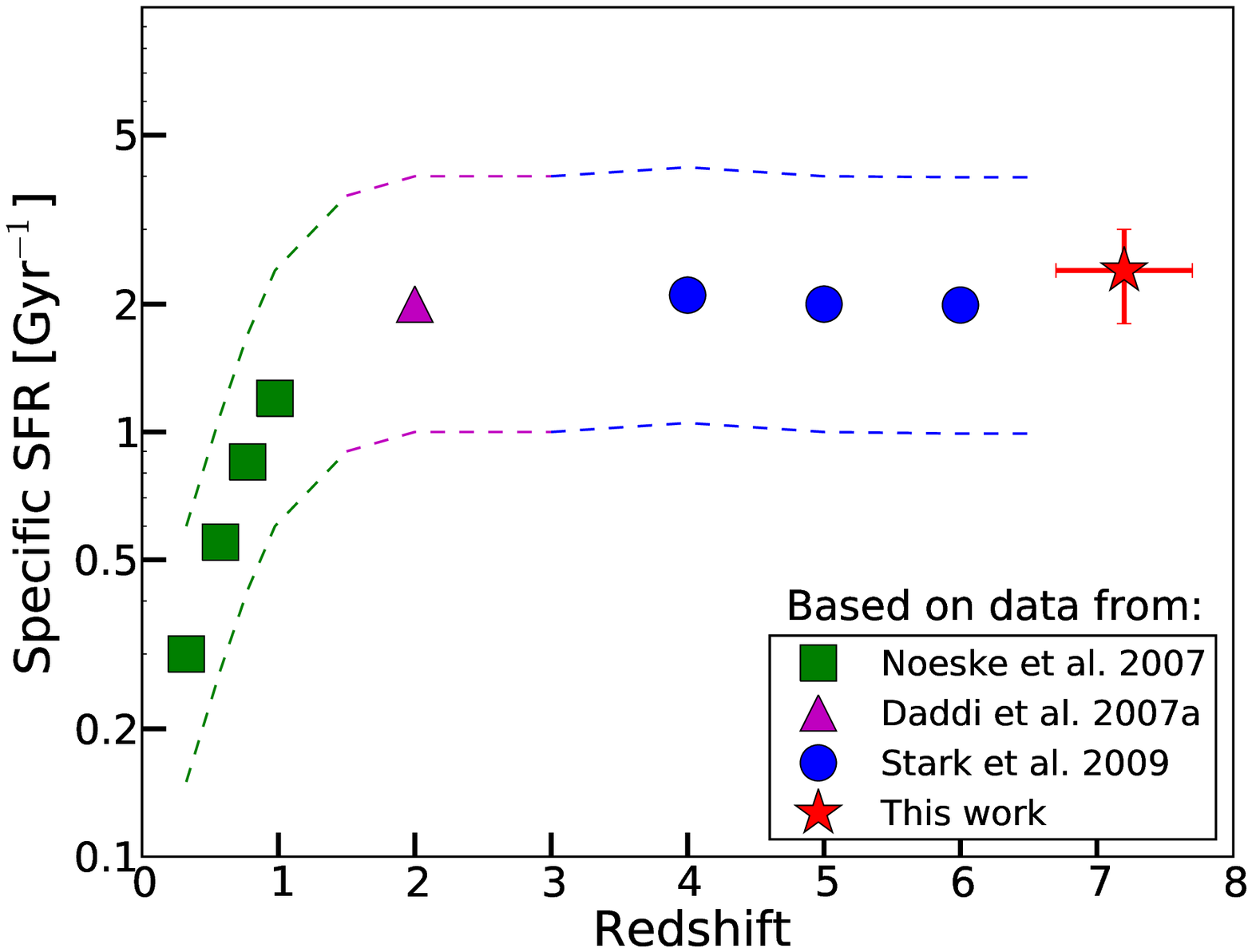}

\caption{The SSFR measured from our data at $z\sim7$ compared to the values 
derived from the data presented by others at a constant stellar mass of 
5$\times10^9 M_\odot$ (corresponding to the median of the present sample). Our 
estimates of the SSFR are based on results from \citet{Noeske2007}, 
\citet{Daddi2007} (in good agreement with \citealt{Papovich2001}), and 
\citet{Stark2009}. We estimate that the typical errors at z$<7$ (dashed lines) 
are $\sim0.3$ dex.  The SSFR  seems to be remarkably constant at 2 Gyr$^{-1}$ 
between $z\sim2-7$ suggesting that the star formation - mass relation does not 
evolve strongly between $z\sim2-7$.  The drop observed at $z<2$, however, 
indicates that some physical process might be inhibiting star formation.  This 
result suggests that star formation in galaxies at $z\gtrsim2$ follows somewhat 
different principles than for galaxies at $z\lesssim2$.}
\label{ssfr} 
\end{figure}
}

\def\figSMD{
\begin{figure}[t]
\centering
\includegraphics[width=0.5\textwidth]{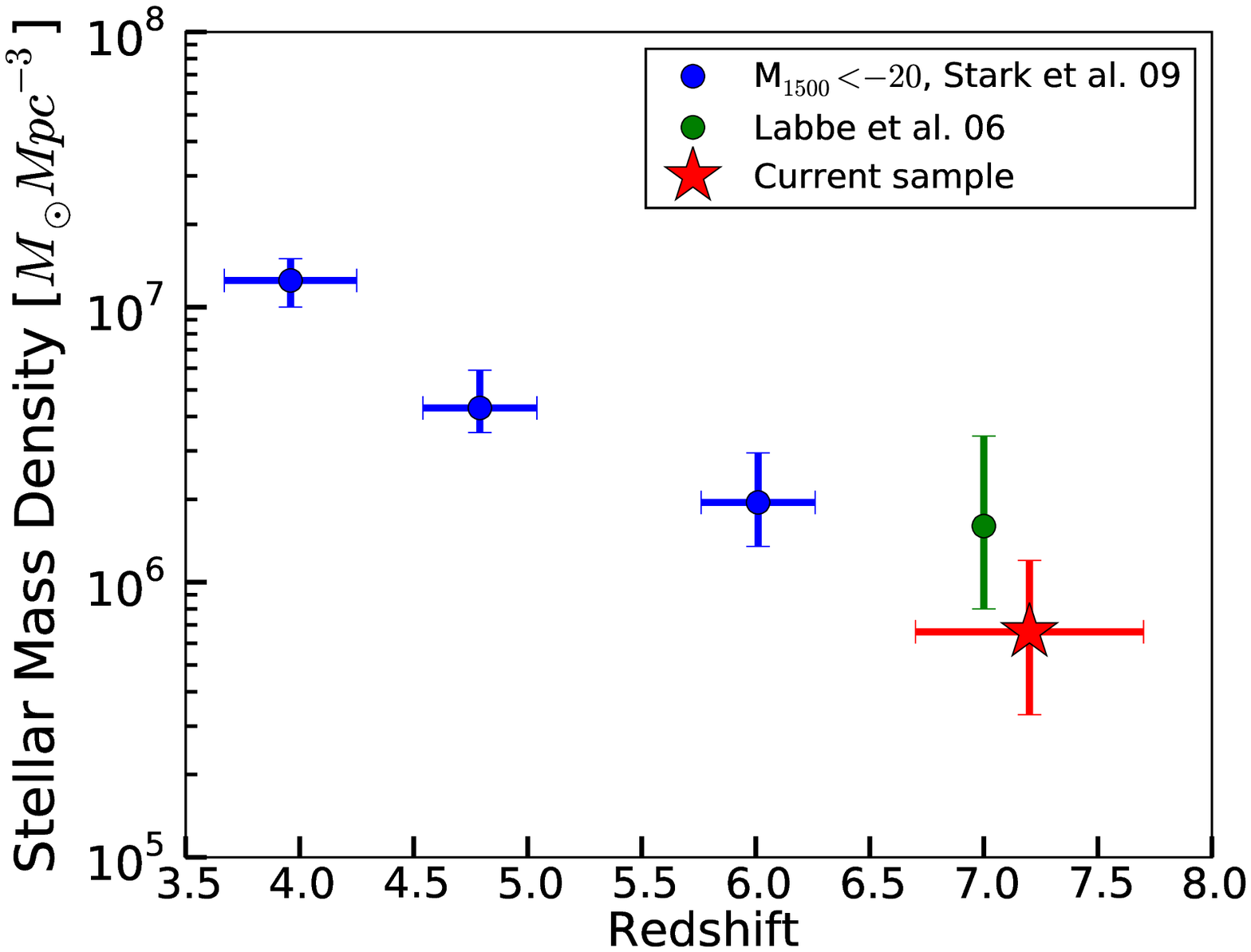}

\caption{The stellar mass density as a function of redshift. Our estimate of the 
stellar mass density at $z\sim7$ is shown with the red star and considers the 
contributions from galaxies with absolute magnitudes $M_{UV,AB}<-20$.  The 
horizontal error bars show the approximate width of our samples in redshift 
space.  Also shown are the stellar mass density determinations of 
\citealt{Stark2009} at $z\sim4$, $z\sim5$, and $z\sim6$ (blue points).  The 
green point at $z\sim7$ is from \citealt{Labbe2006}.  That estimate was derived 
from two $z\sim7$ candidates in the HUDF identified by \citet{Bouwens2004}.  
Those candidates are also included in this study.  The limiting UV luminosity 
probed by our sample is comparable to the points at lower redshift.}
\label{smdz} 
\end{figure}
}

\def\figSFRz{
\begin{figure}[t]
\centering
\includegraphics[width=0.5\textwidth]{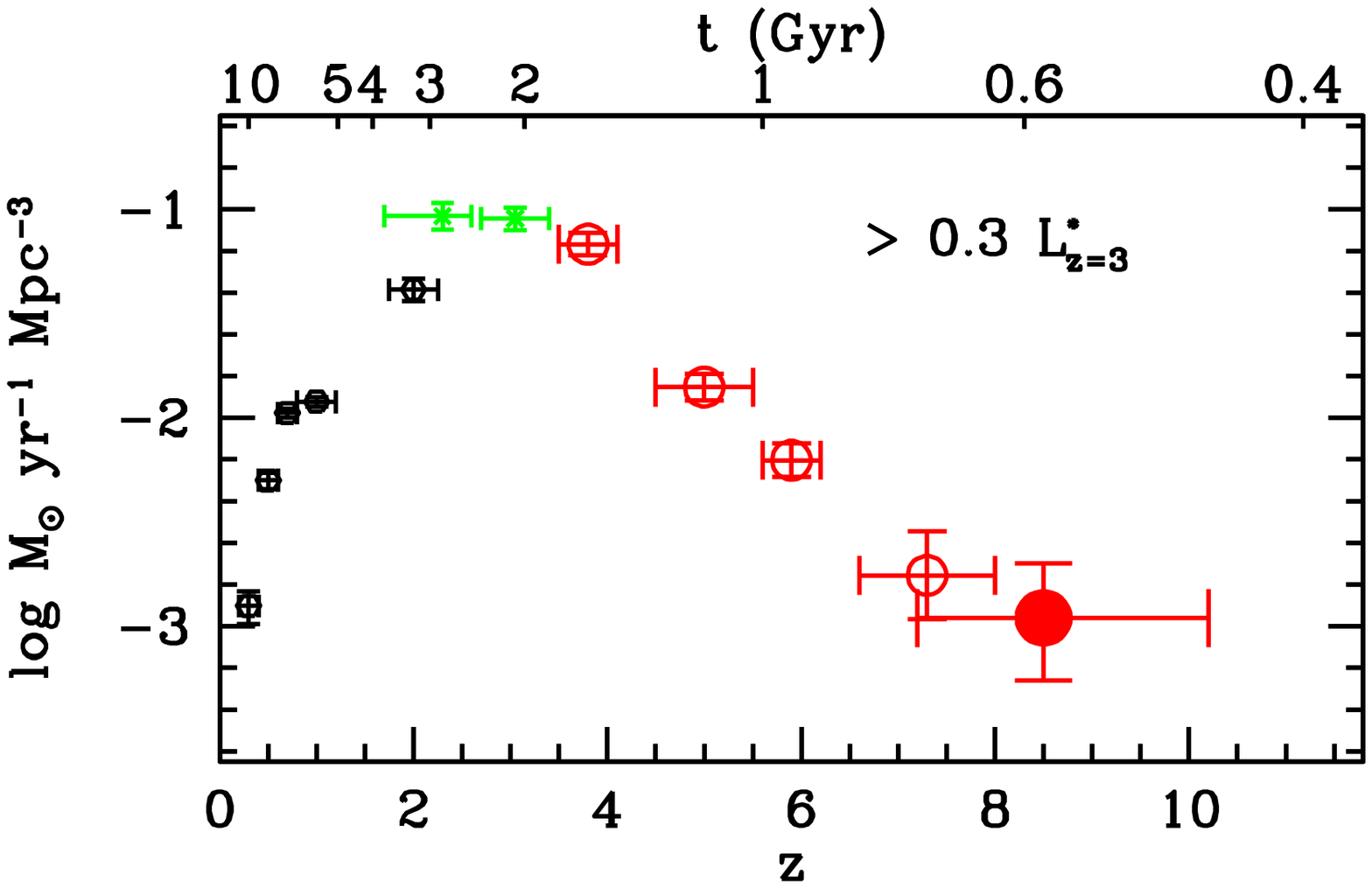}

\caption{The approximate star formation rate density inferred at $z\sim8.5$ 
(solid red circle) by combining the age constraints we have on $z\sim7$ galaxies 
with the stellar mass density we estimate (\S7.3).  For our estimated SFR 
density, we use the stellar mass density we derive at $z\sim7$ divided by the 
average age of our sample.  Included on this figure are the dust-corrected SFR 
density determinations at $z\sim7$ from $z\sim7$ $z$-dropout search 
(\citealt{Bouwens2008}: \textit{open red circle}), the \citet{Bouwens2007}
determination at $z\sim4-6$ (\textit{open red circles}), the \citet{Reddy2009} 
determinations at $z\sim$2-3 (\textit{green crosses}), and the 
\citet{Schiminovich2005} determinations at $z\lesssim2$ (\textit{black 
hexagons}).  Dust corrections are as in \citet{Bouwens2009a}.}
\label{sfz}
\end{figure}
}

\newcommand{\tablewithdata}{
\begin{deluxetable}{lcl}
\tablecaption{Summary of available data. \label{observations}}
\tablehead{
\colhead{Filter} & \colhead{Depth} &\colhead{Reference}\\
&\colhead{[1$\sigma$]}&\\
}
\startdata
\multicolumn{3}{c}{HUDF}\\
ACS - \BB&29.9&[1]\\
ACS - \vv&30.4&[1]\\
ACS - \ii&29.9&[1]\\
ACS - \zz&29.4&[1]\\
NICMOS - \JJ&28.7&[5, 6]\\
NICMOS - \HH&28.3&[5, 6]\\
ISAAC - $K_s$&27.3&[11]\\
IRAC - 3.6\mum&27.5&[15]\\
IRAC - 4.5\mum&26.8&[15]\\
IRAC - 5.8\mum&25.8&[15]\\
IRAC - 8.0\mum&25.7&[15]\\

\tableline
\multicolumn{3}{c}{GOODS South}\\
ACS - \BB&28.8&[2, 3]\\
ACS - \vv&29.1&[2, 3]\\
ACS - \ii&28.5&[2, 3]\\
ACS - \zz&28.3&[2, 3]\\
NICMOS - \JJ&28.0&[7, 8]\\
NICMOS - \HH&27.9&[7, 8]\\
IRAC - 3.6\mum&27.2&[15]\\
IRAC - 4.5\mum&26.4&[15]\\
IRAC - 5.8\mum&25.4&[15]\\
IRAC - 8.0\mum&25.3&[15]\\

\tableline
\multicolumn{3}{c}{GOODS North}\\
ACS - \BB&28.7&[2, 3]\\
ACS - \vv&29.0&[2, 3]\\
ACS - \ii&28.6&[2, 3]\\
ACS - \zz&28.5&[2, 3]\\
NICMOS - \JJ&27.8&[7, 8]\\
NICMOS - \HH&27.7&[7, 8]\\
MOIRCS - $J_s$&27.1\tablenotemark{a}&[12, 13, 14]\\
MOIRCS - $K_s$&27.4\tablenotemark{a}&[12, 13, 14]\\
IRAC - 3.6\mum&27.2&[15]\\
IRAC - 4.5\mum&26.4&[15]\\
IRAC - 5.8\mum&25.4&[15]\\
IRAC - 8.0\mum&25.3&[15]\\

\tableline
\multicolumn{3}{c}{HDFN\tablenotemark{b}}\\
ACS - \BB&29.2&[2, 3]\\
WFPC2 - $B'_{450}$&30.2&[4]\\
ACS - \vv&29.4&[2, 3]\\
WFPC2 - $v'_{606}$&29.0&[4]\\
ACS - \ii&28.6&[2, 3]\\
WFPC2 - $I_{814}$&28.8&[4]\\
ACS - z$_{850}$&28.23&[2, 3]\\
NICMOS - \JJ&27.9&[9, 10]\\
NICMOS - \HH&28.1&[9, 10]\\
MOIRCS - $K_s$&27.6&[12, 13, 14]\\
IRAC - 3.6\mum&26.8&[15]\\
IRAC - 4.5\mum&26.2&[15]\\
IRAC - 5.8\mum&25.8&[15]\\
IRAC - 8.0\mum&25.7&[15]\\

\enddata
\tablecomments{The depths of the data in the different fields. Optical to NIR
were estimated using 0.9\arcs\ diameter apertures and corrected to total
assuming stellar profiles (aperture corrections of 5\% and 20\% for the ACS and
NICMOS data respectively). Apertures of 1.8\arcs\ were dropped randomly in empty
regions of the IRAC images and then the fluxes corrected to total assuming PSF
profiles (aperture correction multiplicative factors of 2.4, 2.7, 3.5, and 3.7
for the 3.6 \mum~, 4.5 \mum~, 5.8 \mum~, and 8.0 \mum~images respectively).}
\tablenotetext{a}{Only source ID \gnsfive\ is covered in this MOIRCS image.}
\tablenotetext{b}{Only source ID \hdfone\ is contained in this set.}
\tablerefs{
(1) \citealt{Beckwith2006};
(2) \citealt{Giavalisco2004};
(3) \citealt{Bouwens2007};
(4) \citealt{Williams1996};
(5) \citealt{Thompson2005};
(6) \citealt{Oesch2009a};
(7) Conselice et al. 2009, in preparation;
(8) Bouwens et al. 2009, in preparation;
(9) \citealt{Thompson1999};
(10) \citealt{Dickinson1999};
(11) \citealt{Labbe2003};
(12) \citealt{Kajisawa2006b};
(13) \citealt{Ouchi2007};
(14) \citealt{Bouwens2008};
(15) Dickinson et al. 2009, in preparation
}
\end{deluxetable}
}

\begin{document}

\title{The Stellar Mass Density  and Specific Star Formation Rates of the
 Universe at $z\sim7$}

\author{Valentino Gonz\'alez\altaffilmark{1}, 
Ivo Labb\'e\altaffilmark{2}, 
Rychard J. Bouwens\altaffilmark{1,3}, 
Garth Illingworth\altaffilmark{1}, 
Marijn Franx\altaffilmark{3}, 
Mariska Kriek\altaffilmark{4}, 
Gabriel B. Brammer\altaffilmark{5}
} 
\altaffiltext{1}{Astronomy Department, University of California, Santa Cruz, CA 
95064}
\altaffiltext{2}{Carnegie Obser vatories, 813 Santa Barbara Street, Pasadena, 
CA 91101}
\altaffiltext{3}{Leiden Observatory, LeidenUniversity , NL-2300 RA Leiden, 
Netherlands}
\altaffiltext{4}{Department of Astrophysical Sciences, Princeton University, 
Princeton, NJ 08544}
\altaffiltext{5}{Department of Astronomy, Yale University, New Haven, CT 06520}

\begin{abstract}
We use a robust sample of 11 $z\sim7$ galaxies (\zz-dropouts) to estimate the
stellar mass density of the universe when it was only $\sim$750 Myr old. We
combine the very deep optical to near-Infrared photometry from the HST ACS and
NICMOS cameras with mid-Infrared Spitzer IRAC imaging available through the
GOODS program. After carefully removing the flux from contaminating foreground
sources we have obtained reliable photometry in the 3.6 $\mu$m and 4.5 $\mu$m
IRAC channels. The spectral shapes of these sources, including their rest frame
optical colors, strongly support their being at z$\sim$7 with a mean photometric
redshift of $\langle z \rangle=$7.2$\pm$0.5. We use \citet{Bruzual2003}
synthetic stellar population models to constrain their stellar masses and star
formation histories. We find stellar masses that range over 0.1 -
12$\times10^9~M_\odot$ and average ages from 20 Myr to up to 425 Myr with a mean of
$\sim 300$~Myr, suggesting that in some of these galaxies most of the stars were
formed at $z>8$ (and probably at $z\gtrsim10$). The best
fits to the observed SEDs are consistent with little or no dust extinction, in
agreement with recent results at $z\sim4-8$. The star formation rates (SFR) are
in the range from 5-20 $M_{\odot}~{\rm yr^{-1}}$. From this sample we measure a stellar
mass density of $6.6_{-3.3}^{+5.4}\times10^5~M_\odot~{\rm Mpc^{-3}}$ to a limit
of $M_{UV,AB}<-20$ (or 0.4$L^*_{z=3}$). Combined with a fiducial lower limit for
their ages (80 Myr) this implies a maximum SFR density of $0.008~M_\odot~{\rm
yr^{-1} Mpc^{-3}}$. This is well below the critical level needed to reionize the
universe at $z\sim8$ using standard assumptions. However,
this result is based on luminous sources ($>$L$^*$) and does not include the
dominant contribution of the fainter galaxies. Strikingly, we find that the
specific SFR is constant from $z\sim7$ to $z\sim2$ but drops substantially at
more recent times.
\end{abstract}

\section{Introduction}

Direct observations of galaxies at high redshift from complete, well-defined
searches place strong constraints on galaxy formation and evolution models.
Extensive studies have been made of galaxies selected by the dropout technique
out to redshift 6 (see e.g., \citealt{Stanway2003, Bunker2004, Yan2004a,
Bouwens2006b, Bouwens2007, McLure2009a}) thanks in part to the capabilities of
the HST ACS and NICMOS cameras. Studies at higher redshifts, however, have been
much more challenging. In particular, one of the key issues has been the
derivation of stellar masses. These masses can provide both strong additional
constraints on formation models and information about the star formation rates
at even earlier times, effectively opening a window towards the earliest phases
of galaxy formation. Deriving masses at higher redshifts does however face some
distinct challenges.

At redshifts $z>4$, ACS and NICMOS only access the rest-frame UV continuum,
which is not a reliable tracer of stellar mass. To obtain better constraints on
this quantity through SED fits it is necessary to extend the observations to the
rest-frame optical. Although Spitzer data presents challenges due to its large
instrumental PSF, these rest-frame optical measurements can be made at high
redshifts with the $3.6~\mu$m and $4.5 ~\mu$m IR channels of the IRAC camera
\citep{Fazio2004}. The IRAC $3.6~\mu$m and $4.5~\mu$m bands probe the rest-frame
optical fluxes around $0.5-0.6~\mu$m of $z\sim5+$ sources and reach to
$\gtrsim$26 AB mag at $5\sigma$ in deep $\gtrsim$23 hr integrations. This is a
remarkable achievement for a 0.8-m telescope. Spitzer will continue to play a
unique role in the determination of fundamental properties like ages and stellar
masses of the earliest galaxies until JWST is launched. These Spitzer and HST
data have permitted estimates of stellar masses for large numbers ($\gtrsim$150)
of $z\sim$5-6 sources \citep{Yan2005, Eyles2005, Egami2005, Yan2006, Stark2009, 
Mobasher2005, Bradley2008, Zheng2009}.
One surprising early finding was the number of quite massive
$\sim10^{10}~M_{\odot}$ galaxies in $z\sim6$ samples (e.g.\citealt{Yan2005, 
Eyles2005}).
It suggested that $z\sim5-6$ galaxies exhibited substantial amounts of star
formation at much higher redshifts and earlier times, well into the epoch of
reionization \citep{Stark2007, Yan2006}.

These analyses have also opened the possibility of looking at the SFR-Mass
relation at higher redshifts, or equivalently, at the specific SFR or SSFR
(i.e., SFR/Mass). The SSFR tells us how fast galaxies are growing with respect
to their current stellar masses. While at low redshifts ($0<z<1$) it has been
shown that galaxies grew faster in earlier times than they do today
\citep{Noeske2007}, it has been suggested that at higher redshift ($4<z<6$) the
SFR-Mass relation remains much more constant \citep{Stark2009}. This transition
in the SSFR is an interesting and important result and needs to be assessed over
a long redshift baseline.

It is imperative then to try to extend these studies to $z\ge7$ and investigate
what are the characteristics of the probable progenitors of these rather massive
$z\sim6$ sources. Doing so, however, has been challenging due to the difficulty
in obtaining deep enough near-IR and optical data to robustly identify $z\sim7$
sources. Such sources only start to become ``common'' at $\gtrsim$26.5 AB mag.
Thus, early selections of $z\sim7$ $z$-dropouts from the HUDF only included a
handful of candidates, and the situation has only improved slowly
\citep{Bouwens2004, Bouwens2008, Oesch2009a}. Consequently, very little has been
published on the stellar masses of $z\sim7$ $z$-dropout galaxies
\citep{Egami2005, Labbe2006}. \citet{Labbe2006} performed stellar population
modeling of four ACS \zz-band dropout sources found in the HUDF, and derived
stellar masses of $\sim10^{9} - 10^{10}~M_{\odot}$, ages of $\sim$100-200 Myr,
and was able to estimate a stellar mass density of $1.6\times10^6~M_\odot$
Mpc$^{-3}$ (to $0.3L^*_{z=3} $).

Fortunately, as a result of continued efforts to select $z\sim7$ galaxies from
the growing quantity of deep near-IR NICMOS data, Bouwens et al.\ (2009, in
preparation) have succeeded in substantially expanding the size of current
$z\sim7$ $z$-dropout selections, and now 14 NICMOS-selected, rather luminous
$z\sim7$ $z$-dropout galaxies are known that are amenable to stellar mass
estimates. The recent advent of WFC3 on HST has expanded the number of known
$z\sim7$ sources and now $\sim$25 sources are known at $z\sim7$
(\citealt{Oesch2009b, Bouwens2009b, McLure2009b, Bunker2009, Castellano2009,
Yan2009, Ouchi2009, Hickey2009, Wilkins2009}). Most of the newly discovered
galaxies, however, are too faint to attempt stellar mass estimations on
individual basis, although valuable information can be obtained from these
samples through stacking analysis (see
\citealt{Labbe2009a}\footnote{The \cite{Labbe2009a} work
focuses exclusively on the ultrafaint, sub-L$^*$ sources. Only sources
\udfthree\ and \udffour\ from this work (our faintest \zdrops\ in the UDF) are
also included in \cite{Labbe2009a}.}). Galaxies in the Bouwens et al.\ (2009,
in preparation) sample extend from $\sim$25.5 AB mag to $\sim$27.8 AB mag and
are found in the HUDF, and in and around the two wide-area GOODS fields. These
candidates also possess very deep $\gtrsim$23.3 hr coverage with IRAC from the
Spitzer GOODS program and present us with a unique opportunity to better
understand what the typical properties (ages, stellar masses) of $z\sim7$
$z$-dropouts are.

Here we take advantage of the larger sample of 14 \zz-dropout sources identified
by Bouwens et al.\ (2009, in preparation) to estimate the typical properties of
$z\sim7$ galaxies. The much larger size of current samples allows us to get a
better handle on the typical properties of $z\sim7$ galaxies than was possible
from the smaller and brighter samples previously available \citep{Labbe2006}.
For example, sizeable variations in the M/L ratios of individual galaxies can
considerably skew the averages for the population as a whole. The present sample
also include galaxies from two independent lines of sight, i.e., the HDF-North
GOODS and CDF-South GOODS, so the results should be much less impacted by cosmic
variance and more representative of the cosmic average. In estimating the
rest-frame optical fluxes for our $z\sim7$ candidates from the IRAC, we will
take advantage of the well-tested deblending process described by
\citet{Labbe2006} (see also \citealt{Wuyts2007}) which enables us to estimate
fluxes when there is moderate overlap with nearby sources. This also allows for
a larger sample, since the well-known problems with blending and confusion in
IRAC data have limited previous studies.

We provide a brief outline for this paper here. We present the sample selection
and observational data we have used in this work in \S2 and \S3 respectively. In
\S4, we will describe the photometry of the sample with particular emphasis in
the deblended photometry from the IRAC channels. We devote \S5 to a discussion
of the photometric redshifts obtained and possible contamination and in \S6 we
present the procedure and results in the process of fitting synthetic stellar
populations to the observed SEDs (including our SSFR results). \S7 presents our
estimated stellar mass density (SMD) at $z\sim7$ and confidence intervals. We
compare our results to previous work in \S8 and present a discussion of them in
\S9. We summarize our conclusions in \S10. All magnitudes quoted in the paper
are in the AB system \citep{Oke1983}. We have used cosmological parameters
$\Omega_{\Lambda}=0.7, \Omega_{M}=0.3$ and $H_0= 70~{\rm kms~s^{-1}~Mpc^{-1}} $
to facilitate comparison to previous works.

\section{Sample Selection}

The present selection of $z\sim7$ $z$-dropout candidates is based upon $\sim80$
arcmin$^2$ of very deep optical ACS, near-IR NICMOS, and IRAC data available
over and around the two GOODS fields -- including the HUDF and HDF-North (see
\S3 for a more detailed description of these data). This selection is described
in Bouwens et al.\ (2009, in preparation).

\tablewithdata

Candidates $z\sim7$ galaxies were required to satisfy a
two-color Lyman-Break Galaxy (LBG) criterion adapted to $z\sim7$ -- showing a
strong $z-J$ break but possessing a blue $J-H$ color (redward of the break). In
detail, the color criterion used is:
$$(z_{850}-J_{110})_{AB}>0.8$$
$$\wedge ~(z_{850}-J_{110})_{AB}>0.8+0.4(J_{110}-H_{160})_{AB}$$
where $\wedge$ represents the logical \textbf{AND} operation and where the
colors were measured in a $\sim$0.4$\arcsec$-diameter aperture using
$SExtractor$ in double image mode with the \HH~band image being the detection
image. Candidate $z\gtrsim7$ sources were also required to be undetected to less
than 2$\sigma$ in the optical ACS bands (\BB, $V_{606}$, and \ii). Sources were
also removed from the sample if they were detected at 1.5$\sigma$ in two or more
of the optical bands. Only detections of at least 5$\sigma$ in the
$H_{160}$-band images (measured in apertures of 0.6\arcs~in diameter) were
considered to ensure that our candidates corresponded to real sources. The
candidates --with GNS IDs-- were also required to have have ($H_{160}-5.8~\mu$m)
colors bluer than 2.5 (5.8~$\mu$m photometry measured in 2\arcsec\-diameter
apertures). This provided added confidence that the sources were not
low-redshift interlopers.

By applying the above criteria to $\sim$80 arcmin$^2$ of deep NICMOS data,
Bouwens et al.\ (2009, in preparation) identified 14 $z\sim7$ $z$-dropout
candidates (see Table~\ref{photometry}). The candidates range in $H_{160,AB}$
magnitude from 25.5 to 27.6 AB mag and are typically only marginally resolved at
NICMOS/NIC3 resolution (FWHM of the PSF is $\sim$ 0.37$\arcsec$).
Contamination rates for the sample are determined by running
a number of photometric scattering simulations on a fake sample of low redshift
sources. This sample is constructed to match the color distribution of observed
galaxies in the $24.5<$\HH$<26$ range. We have added noise to the fluxes in each
individual band according to the depths of the different fields. By applying the
selection criteria previously described, we characterize when and with what
frequency low-redshift contaminants enter our selection. A more detailed
discussion of the procedure is given in \citet {Bouwens2008} and Bouwens et al.\
(2009, in preparation). Through this technique we find that the contamination
for the sample (by lower redshift galaxies, time-variable sources, low-mass
stars, for example) is expected to be just $\sim$10\% of the sample for sources
found over the HUDF and HUDF05 fields, and only $\sim$20\% for sources within
the GOODS fields.

We only use those 11 $z$-dropout candidates from the Bouwens et al.\ (2009, in
preparation) selection that have deep (23+ hr) IRAC data in both 3.6 and 4.5
$\mu$m and which were in the Bouwens et al.\ (2009, in prep) $z$-dropout sample
as of June 2009.

\begin{deluxetable*}{llccccccccc}
\tabletypesize{\footnotesize}
\tablecaption{Summary of Photometry. \label{photometry}}
\centering
\tablewidth{0pt}
\tablehead{
\colhead{ID}&\colhead{Field}&\colhead{\BB}&\colhead{\vv}&
\colhead{\ii}&\colhead{\zz}&\colhead{\JJ}&\colhead{\HH}&
\colhead{$K_s\tablenotemark{a}$}&\colhead{3.6\mum}&\colhead{4.5\mum}\\
}
\startdata
\udfone&HUDF&$>$29.9&$>$30.4&$>$29.9&27.8$\pm$0.2&26.5$\pm$0.2&26.1$\pm
$0.2&
25.8$\pm$0.2&25.2$\pm$0.1&25.9$\pm$0.4\\
\udftwo&HUDF&$>$29.9&$>$30.4&$>$29.8&$>$29.2&26.7$\pm$0.2&26.8$\pm$0.3&
26.6$\pm$0.6&26.2$\pm$0.3&26.6$\pm$0.9\\
\udfthree&HUDF&$>$30.0&$>$30.6&$>$30.0&29.1$\pm$0.8&27.7$\pm$0.4&26.9$\pm
$0.3&
$>$27.4&$>$27.5&$>$26.8\\
\udffour&HUDF&$>$30.0&$>$30.4&$>$30.0&$>$29.4&27.6$\pm$0.3&26.9$\pm$0.3&
$>$27.1&
26.5$\pm$0.4&26.7$\pm$1.0\\
\gnsone&GOODS-S&$>$28.6&$>$29.0&$>$28.3&$>$27.9&26.5$\pm$0.3&26.2$\pm
$0.2&\nodata&
24.9$\pm$0.1&25.4$\pm$0.4\\
\gnstwo&GOODS-S&$>$28.9&29.0$\pm$1.0&$>$28.4&$>$28.4&27.0$\pm$0.4&26.7$
\pm$0.3&
\nodata&26.4$\pm$0.6$^{*}$&26.2$\pm$0.9\\
\gnsthree&GOODS-S&$>$29.1&$>$29.4&$>$29.0&$>$28.7&27.3$\pm$0.7&26.7$\pm
$0.3&\nodata&
25.7$\pm$0.3&25.9$\pm$0.7\\
\gnsfour&GOODS-N&$>$28.5&$>$28.8&$>$28.4&$>$28.4&26.7$\pm$0.4&26.3$\pm
$0.3&\nodata&
25.2$\pm$0.2&25.4$\pm$0.4\\
\gnsfive\tablenotemark{b}&GOODS-N&$>$28.8&$>$29.1&$>$28.7&28.1$\pm
$0.7&26.2$\pm$0.3
&25.3$\pm$0.1&25.4$\pm$0.2&24.6$\pm$0.1$^{*}$&25.4$\pm$0.3\\
\cdfone&GOODS-S&$>$28.7&$>$29.0&$>$28.6&$>$28.4&26.8$\pm$0.4&26.4$\pm
$0.3&\nodata&
26.1$\pm$0.3&26.0$\pm$0.5\\
\hdfone\tablenotemark{c}&HDFN&$>$29.2&$>$29.4&$>$28.6&26.5$\pm$0.2&26.2$
\pm$0.2
&25.9$\pm$0.2&26.4$\pm$0.4&25.2$\pm$0.3$^{*}$&25.7$\pm$0.7\\
\tableline\\
\multicolumn{2}{l}{Mean SED\tablenotemark{d}}&$>$30.3&$>$30.6&$>$30.1&28.41$
\pm$0.29&26.70$\pm$0.09&
26.27$\pm$0.07&26.68$\pm$0.30&25.58$\pm$0.07&25.91$\pm$0.18\\
\enddata
\tablecomments{Magnitudes are total and in the AB system.  Optical to near 
infrared photometry measured in 0.9\arcs\ diameter apertures with aperture 
corrections of 5\% - 20\% (derived assuming stellar profiles).  Mid infrared 
IRAC photometry performed on ``cleaned'' images with 1.8\arcs\ aperture 
diameters.  Aperture correction factors in this case are 2.4 and 2.7 in the 3.6 
and  4.5 \mum\ channels respectively.  Upper limits and error bars are 1 
$\sigma$.}
\tablenotetext{a}{K-band from either MOIRCS or ISAAC depending on the Field, see 
Table \ref{observations}.}
\tablenotetext{b}{MOIRCS J$_s$ band imaging is deep enough at the location of 
this source.  We measure J$_s$=25.6$\pm$0.3.}
\tablenotetext{c}{There are additional WFPC2 optical imaging constraints for 
\hdfone:  B$>30.2$, V$>29.0$, $i>$28.8.}
\tablenotetext{d}{This mean SED was constructed rescaling all the SEDs so that
the \HH-band fluxes coincide with the mean value and then averaging all the
other bands. The 3.6\mum\ IRAC fluxes from sources \gnstwo, \gnsfive, and
\hdfone\ were excluded from the average. See section \S6.2 and figure
\ref{meansed}.} \tablenotetext{*}{After subtracting a bright nearby neighboring
from the 3.6\mum\ image of this source, sizable residuals remain, and so the
quoted 3.6\mum\ flux measurements for these sources may suffer from systematic
errors.}
\end{deluxetable*}

\section{Observational Data}

The very deep optical, near-IR, and mid-IR IRAC data available for our $z\sim$7
$z$-dropout candidates permit us to study the properties of these sources in
great detail. A summary of the available imaging data for the candidates we have
in our various search fields is given in Table~\ref{observations}.

The deep near-IR data we have available for our candidates
at $\sim$1.1~$\mu$m and $\sim$1.6~$\mu$m comes from NICMOS. For the candidates
within the GOODS fields, the NICMOS data reach to depths of $\sim$26.8 AB mag
and $\sim$26.7 AB mag ($5\sigma$, aperture flux within a 0.6\arcs$-$diameter,
used for detection ) in the \JJ\ and \HH\ bands, respectively. For purposes of
the SED fitting we use a larger (0.9\arcsec-diameter) aperture to minimize
differences in the aperture corrections among the optical, NIR and mid-IR
images. The 1$\sigma$ limits relevant for the SED fitting are $\sim27.8$ and
$\sim27.9$ AB mag (\JJ\ and \HH\ bands, corrected to total fluxes using aperture
corrections of $\sim20\%$ derived from stellar profiles). The near-IR data over
the HUDF reaches some $\sim$0.5 AB mag deeper. The FWHM of the NICMOS PSF is
0.34\arcsec~and 0.37\arcsec~in the \JJ\ and \HH\ bands, respectively.

We also have very deep $\sim$2.2~$\mu$m K-band data available for our $z\sim7$
$z$-dropout candidates over the HUDF and in the central region of the HDF-North.
These data are particularly valuable for providing a constraint on the
UV-continuum slopes. The $K$-band data over the HUDF correspond to the 40 hr
integration in the best seeing conditions in the K$_{s}$-band filter at ISAAC
(VLT) and PANIC (Magellan) \citep{Labbe2006}. We estimate 1$\sigma$ depths of
$\simeq27.1-27.4$ AB magnitudes in 0.9\arcsec~diameter apertures. The other two
sources are in the HDFN and their K-band data come from the deep Subaru GTO
MOIRCS imaging campaign (\citealt{Kajisawa2006b, Ouchi2007}: see
\citealt{Bouwens2008} for a description of our reductions). We have only made
use of the deepest GTO pointing which reaches down to 25.4 total AB magnitudes
at 5$\sigma$ (0.9\arcsec~diameter apertures). The FWHM for the $K$-band PSF is
$\sim$0.5\arcsec.

Deep optical \BB, \vv, \ii, and \zz\ observations are available for our
candidates with ACS and typically reach to depths of $\sim30$ mag in the HUDF
and $\sim$1.5 mag shallower in the rest of our fields. For the single
$z$-dropout candidate in the HDF-North WFPC2 field, we have the very deep
($\gtrsim28$ mag at $5\sigma$) $B_{450}, V_{606}$, and $I_{814}$ WFPC2
observations -- which permit us to set very strong constraints on the strength
of the Lyman Break.

The Spitzer IRAC imaging data from the GOODS program (Dickinson et al.\ 2009, in
preparation) provide us with deep rest-frame optical coverage on our $z\sim7$
$z$-dropout candidates -- which is critical for estimates of the stellar mass in
these sources. Two exposures of $\sim$23-hr each were taken in two different
epochs with the IRAC camera -- rotated by $180\deg$ -- and overlapping in the
center of the GOODS field. The region of overlap contains the HUDF in the GOODS
South. We find AB magnitude detection limits for point
sources of 27.4, 26.6, 25.4, and 25.3 for the 3.6, 4.5, 5.8, and 8.0 $\mu$m
channels respectively (1 $\sigma$, measured on apertures of 1.8\arcsec~in
diameter and corrected to total flux assuming stellar profiles with aperture
corrections of 2.4, 2.7, 3.5, 3.7 respectively -- multiplicative factors).
These limiting depths (for single epoch $\sim$23.3 hr IRAC observations) were
estimated by dropping apertures at random empty regions of the sky and measuring
the flux variations. A good summary of the IRAC observations is provided in
\citet{Labbe2006} and \citet{Stark2007}, for example. In this work we make use
of the reductions of Data Release (DR) 3 of epoch 1 observations and DR2 of
epoch 2 of the GOODS-S field. In the case of the GOODS-N field we make use of
the reductions of Data Release (DR) 2 of both epochs.

\section{Photometry}

\textit{Optical/near-IR Photometry:} Optical to NIR fluxes were measured in
standard 0.9\arcsec-diameter circular apertures. We corrected these measured
fluxes for the missing light outside these apertures assuming stellar profiles.
These latter corrections increased the measured flux by 5\% - 20\% depending on
the band. All the sources in our sample were sufficiently separated from their
neighbors that we could use this simple approach, except for the one in the HDFN
field where there was a faint but very close (almost overlapping) neighbor. This
made it impossible to use the standard circular aperture to measure the flux of
\hdfone~ (ID from Bouwens et al. 2009, in preparation). To ensure our
measurement of its flux was not contaminated by the flux of this neighbor, we
fitted a PSF profile to the neighbor using GALFIT \citep{Peng2002} and
subtracted it from the image before measuring the flux of this $z$-dropout.

\textit{IRAC Photometry:} One of the biggest challenges in estimating the masses
of our $z\sim7$ candidates is the acquisition of reliable mid-IR fluxes for
these candidates from the available IRAC data. The extremely broad PSF of this
instrument and large pixel sizes make the images extremely crowded and so fluxes
from neighboring sources spill over onto each other. To overcome this issue, a
wide variety of different approaches have been developed, almost all of which
involve modeling the IRAC image with a number of smoothed sources of varying
flux. In the most common cases, the model light profiles are theoretical like
the ones produced by GALFIT. The use of these models can result in systematic
errors in the photometry if the sources have irregular or clumpy spatial
profiles.

We have used here the technique described in \citet{Labbe2006}, which consists
in the creation of an empirical light profile based in the higher resolution
NICMOS images. This technique uses a segmentation map created by $SExtractor$
\citep{Bertin1996} to define the boundaries of each source in the area to be
cleaned (we use a 2$\sigma$ threshold to ensure all the
possibly relevant neighbors are fitted) and use the light profiles of the
sources within those boundaries as the empirical light profiles (assuming a
similarity between the profiles at both 1.6~$\mu$m and $3.6~\mu$m). The
individual profiles are then convolved with a carefully constructed kernel
(based on the instrumental PSFs) to simulate how they would look like in the
IRAC images (modulo a normalization factor). Finally we fit for the total flux
of each neighbor and subtract them off the image. Instead of using the
$z$-dropout flux measurement determined from these fits, we subtract off the
flux from the neighbors and perform standard aperture photometry in relatively
small apertures on the ``cleaned'' image. We find that the optimal aperture
diameter for maximizing the S/N of our flux measurements (assuming point
sources) is $\sim$1.8\arcsec~in all the channels. The aperture corrections are
2.4, 2.7, 3.5, and 3.7 (multiplicative factors) in the 3.6, 4.5, 5.8, and 8.0
$\mu$m channels respectively. The errors in the measured flux include both the
typical variations on the sky brightness, and the uncertainty in the flux
removed from the aperture.

A simple inspection of our images shows that 5 of the 11 $z\sim7$ $z$-dropout
candidates in our sample (i.e., \gnstwo, \gnsfive, \udfthree, \udffour, \hdfone)
are severely blended with bright foreground sources. The other six sources are
somewhat more isolated but with the size of the instrumental PSF it is obvious
that flux from neighboring sources will spill over onto these candidates (often
contributing $\gtrsim$20\% of the light within the $1.8\arcsec$-diameter
aperture centered on our candidates). From the 5 severely blended sources we are
able to satisfactorily ``clean'' the images in at least 2 cases (we will discuss
the other three cases \gnstwo, \gnsfive\ and \hdfone\ in the next paragraph). We
could check the consistency of the method for 5 of the sources that were imaged
in both epochs. As was mentioned before, there is a 180 $\deg$ rotation in the
IRAC camera (and thus in the asymmetric instrumental PSF) between the two
epochs, which makes the light profile models (for the neighboring sources)
almost independent. Obtaining consistent fluxes in these two images is a good
indication of the reliability of the method. For the sources where IRAC
observations are from $>$1 epoch (and consistent within the measurement errors),
we average the fluxes and combine the errors accordingly which reduces them by a
factor $\sqrt{2}$.

\figstamps

In the cases of \gnstwo\ and \gnsfive\ strong color gradients in the closest
neighbor cause the model light profiles to be inadequate and so large residuals
are evident (after subtracting the flux from the neighbors). \gnsfive\ was
imaged in both epochs and we find a difference of $2\sigma$ between the
measurements in the 3.6 $\mu$m channel but a much cleaner residual (and better
agreement) in the 4.5 $\mu$m image. A similar discrepancy was found for the 3.6
$\mu$m flux measurement for \hdfone. In this case the poor subtraction is not
due to strong color gradients but to the extreme proximity of its neighbor. In
all three cases we adopt the single epoch uncertainty. Given the small size of
our present sample of $z$-dropouts, we chose to keep these three sources in the
sample -- keeping in mind the caveat that the flux measurements for these
sources could possess large systematic errors. In summary,
we are able to perform reliable cleaned photometry on 8 out of the 11 sources (73\%).

We obtain $>2\sigma$ detections for 9 of the 11
sources in the 3.6 \mum\ image and for 5 sources in the 4.5 \mum\ image. The two
sources with quite marginal ($<2\sigma$) detections in both IRAC images are
\udfthree\ and \gnstwo. A simple stacking of the images of these two sources
(adding both sources in both IRAC images) shows a significant detection, which
provides evidence for the reality of the sources.

Optical to mid-Infrared image stamps ($\sim4\arcsec\times4\arcsec$) for all the
sources are presented in figure \ref{stamps}. The 5.8 $\mu$m and 8.0 $\mu$m
channels have been omitted because none of the sources are detected in those
bands (as expected). The two epochs have been coadded when available. The
measured magnitudes are summarized in table \ref{photometry}.

\section{Photometric Redshifts}

Perhaps the most fundamental quantity that we can estimate for galaxies in our
\zz-dropout sample is their redshift. We first explore the probable redshifts of
our sources using the photometric redshift code EAZY \citep{Brammer2008}. The
code works by comparing the observed photometry with that predicted on the basis
of the specific SED templates. We use the default template set, which was
derived from the Pegase population synthesis models \citep{Fioc1997} and
optimized to reproduce the properties of galaxies in the range $0<z<4$.
Comparisons of the photometric redshift estimates output from EAZY with some of
the deepest spectroscopic surveys available show minimal systematic errors and a
scatter in $z_{spec}-z_{phot}$ of $\sigma$=0.034 over the range $0<z<4$ (see
table 1 in \citealt{Brammer2008}). The few sources at higher redshifts
($z\sim6$) with available spectroscopy are also in good agreement with its
photometric redshift estimates. One relevant advantage over other photometric
redshift codes is that EAZY works with fluxes instead of magnitudes and
naturally handles negative measurements which are common in our optical bands.

\fighist

We use the available flux measurements in the \BB, $V_{606}$, \ii, \zz, \JJ,
\HH, 3.6 $\mu$m, 4.5 $\mu$m bands, and also the $B_{450}$, $V_{606}$, $I_{814}$,
$K$ bands if available for our photometric redshift estimates. We do not include
the IRAC 5.8 $\mu$m and 8.0 $\mu$m flux measurements of the sources (consistent
with no detection) in the comparison -- since they do not help us to
meaningfully discriminate between the competing redshift solutions.
We restrict the redshift range of our fits to $z\sim4-11$
and adopt no redshift prior. Solutions at $z\sim1.5$ are also possible for most
of the sources but at lower probability for all of them. However, quantifying
this is challenging given our poor knowledge of the demographics of galaxies at
both $z\sim1.5$ and $z\sim7$. Furthermore, the existence of synthetic solutions
at any of these redshifts does not necessarily imply the existence of real
galaxies with the observed properties. We have provided an independent (short)
description of the estimate of the fraction of low-redshift contaminants in \S2.
A more detailed discussion is provided in the Bouwens et al.\ (2009, in
preparation) study where the sample is described (but see also
\citealt{Bouwens2008} which describes the estimate for a similar sample). The
estimated contamination fraction for this sample is $\sim10\%$ and $\sim20\%$
for sources in the HUDF and outside the HUDF, respectively.

We explore the solutions in the range $4<z<11$ in a grid with steps of 0.01. At
each step there is a $\chi^2(z)$ value associated with the best solution that is
used to create a probability function $p(z)\propto \exp[-\chi^2(z)/2]$. The
redshift is estimated marginalizing over this probability and probability
contours are used to determine the 68\% confidence intervals. Because of the
shape of this function at redshifts $z>7$, the estimated redshifts are usually
located somewhat above the absolute minimum $\chi^2$. The typical uncertainties
we estimate for individual sources are in the order of 0.6. Finally, the
redshift distribution obtained from EAZY is $z =7.2\pm0.5$, similar to what
would be expected based on the shape of the ACS \zz\ passband and the color cuts
imposed in the sample(see figure~\ref{zdist}).

\section{Stellar Population modeling}

We can make sense of the present photometric selection of $z$-dropouts in terms
of their \textit{intrinsic} properties -- like age, stellar mass, or dust
extinction -- by modeling the spectral energy distributions observed for
individual sources. Such modeling is now ubiquitous in the literature
\citep{Sawicki1998, Brinchmann2000, Papovich2001, Labbe2007}, and has proven
quite powerful in the interpretation of distant galaxies -- both given the
abundance (and quality) of photometric data and the plausibility of the
photometric estimates. Indeed, studies have found reasonable agreement between
stellar masses determined from such modeling and those determined from the
dynamics \citep{Erb2006b}.

Here we model the stellar populations with the \cite{Bruzual2003} (BC03)
spectral synthesis libraries. Over the wavelength range covered by the present
study, these libraries show almost no difference with respect to the newest
Charlot \& Bruzual (2007) (CB07, e.g. \citealt{Stark2009, Labbe2009a}) and have
been shown to be in reasonable agreement with other libraries (e.g.,
\citealt{Maraston2005}). We use a \citet{Salpeter1955} Initial Mass Function
(IMF) between 0.1 and 100 M$_\odot $ and sub-solar metallicity (0.2 Z$_\odot$)
so that comparisons to existing works are more straightforward.
Our selection of sub-solar metallicity models is based on the observed trends at 
high redshifts (e.g. Maiolino et al 2008) as well as  on the direct observations
of extremely blue UV slopes of the most recently found $z\sim7$ sources 
(Bouwens et al. 2010b, Labbe et al. 2010).  However, since
other IMFs (or metallicity models) fit our observations just as well, we will
remark on how the results change if we adopt a different IMF or
metallicity. For simplicity, we assume a constant star formation (CSF) rate when
modeling the star formation histories of the galaxies in our sample. This
assumption seems preferable to an exponentially-decaying SF history (which
implies increasing SFRs with redshift), as there is currently no evidence that
UV-selected samples form stars at a faster rate at earlier epochs (e.g.,
\citealt{Papovich2001}, \citealt{Stark2009}, and others).

\figseds

\begin{deluxetable*}{lcccccccccr}
\tablecolumns{11}
\tablewidth{0pt}
\tablecaption{Summary of BC03 Models fit parameters. 0.2 $Z_\odot$, Salpeter
IMF, constant SFR, $A_{v}=0$.\label{fits}}
\tabletypesize{\footnotesize}
\centering
\tablehead{
\colhead{ID}&\colhead{$z_{phot}$}&\colhead{Mass}&\colhead{Age$_w$\tablenotemark
{a}}&\colhead{SFR$_{L_{1500}}$}&\colhead{SFR}&\colhead{SSFR\tablenotemark{b}}&
\colhead{L$_{UV}$}&\colhead{U-V}&\colhead{$H_{160}-3.6$}&\colhead{$\chi^2_{red}$
}\\ 
&&\colhead{$[10^9M_{\odot}]$}&\colhead{[Myr]}&\colhead{$[M_{\odot}yr^{-1}]$}&
\colhead{$[M_{\odot}yr^{-1}]$}&\colhead{[Gyr$^{-1}$]}&\colhead{$[10^{10}L_{\odot
}]$}&&&\\
}
\startdata
\udfone&$6.9^{+0.1}_{-0.1}$&$6.6^{+0.3}_{-0.9}$&$379$&13.2&$10.5^{+0.7}_{-0.7}$&
$2.0^{+0.3}_{-0.1}$&5.4&0.5&0.9&1.7\\ 
\udftwo&$7.3^{+0.4}_{-0.3}$&$2.2^{+2.0}_{-1.5}$&$173$&9.0&$7.4^{+2.4}_{-1.4}$&
$4.1^{+8.9}_{-1.9}$&3.7&0.1&0.6&0.4\\ 
\udfthree&$7.1^{+2.0}_{-0.5}$&$0.2^{+1.4}_{-0.1}$&$ 19$&4.8&$5.2^{+5.0}_{-2.4}$&
$25.6^{+12.3}_{-22.6}$&2.0&-0.7&-0.8&0.3\\ 
\udffour&$7.9^{+0.8}_{-0.6}$&$2.8^{+0.5}_{-2.1}$&$315$&6.8&$5.4^{+2.0}_{-1.2}$&
$2.4^{+6.9}_{-0.4}$&2.8&0.4&0.4&0.9\\ 
\gnsone&$7.2^{+0.2}_{-0.2}$&$7.6^{+0.4}_{-0.5}$&$362$&15.8&$12.6^{+1.2}_{-1.1}$&
$2.1^{+0.1}_{-0.1}$&6.5&0.5&1.3&1.5\\ 
\gnstwo&$7.1^{+1.5}_{-0.6}$&$2.5^{+1.8}_{-2.2}$&$251$&7.3&$5.9^{+6.4}_{-1.6}$&
$3.0^{+24.7}_{-1.3}$&3.0&0.3&0.3&0.4\\ 
\gnsthree&$7.3^{+0.9}_{-0.4}$&$4.2^{+0.4}_{-1.5}$&$354$&8.8&$7.1^{+2.0}_{-1.1}$&
$2.1^{+1.2}_{-0.2}$&3.6&0.4&1.0&0.3\\ 
\gnsfour&$7.2^{+0.4}_{-0.2}$&$6.8^{+0.3}_{-0.7}$&$362$&13.9&$11.2^{+1.4}_{-1.2}$
&$2.1^{+0.3}_{-0.1}$&5.7&0.5&1.1&0.8\\ 
\gnsfive&$7.3^{+0.2}_{-0.2}$&$12.3^{+0.3}_{-2.1}$&$354$&26.1&$20.9^{+1.5}_{-1.4}
$&$2.1^{+0.4}_{-0.0}$&10.7&0.4&0.7&2.3\\ 
\cdfone&$7.1^{+1.5}_{-0.5}$&$3.5^{+2.0}_{-2.3}$&$281$&9.2&$7.4^{+5.2}_{-1.8}$&
$2.7^{+5.2}_{-1.0}$&3.8&0.3&0.3&0.6\\ 
\hdfone&$6.3^{+0.2}_{-0.2}$&$6.9^{+0.3}_{-3.8}$&$426$&12.4&$9.8^{+1.7}_{-0.9}$&
$1.8^{+2.2}_{-0.1}$&5.1&0.5&0.7&0.8\\ 
\tableline\\
Mean SED\tablenotemark{c}&$7.3^{+0.1}_{-0.0}$&$6.3^{+0.1}_{-0.1}$&$354$&13.4&
$10.7^{+0.5}_{-0.2}$&$2.1^{+0.0}_{-0.0}$&5.5&0.4&0.7&3.3\\ 
\enddata
\tablecomments{Best fit parameters, 68\% confidence intervals and corresponding
$\chi^2$ for \citet{Bruzual2003} fits with subsolar metallicity (0.2Z$_\odot$)
and Salpeter IMF between 0.1 and 100 $M_\odot$. We obtain a redshift for the
sample of 7.2$\pm$0.5. The models suggest that in some of these galaxies most of
the stars were born at considerably earlier times ($z\lesssim10$), well into the
epoch of reionization. The masses of these sources range 0.2-12$\times
10^9~M_{\odot}$. We have restricted the dust extinction to zero, consistent with
the trend observed at these high redshits. SFRs from the models are consistent
with the ones derived from the extrapolated $L_{1500}$ using the usual
\citealt{Madau1998} formula. These SFRs are somewhat high,
$\sim10M_{\odot}yr^{-1}$ typically. The mean SSFRs of the sample is 2.4$\pm$0.6
Gyr$^{-1}$, with an outlier corresponding to the youngest model. The estimated
best fit ages do not change considerably if we consider either Chabrier IMF or
solar metallicity models. As expected for a Chabrier IMF, the masses derived are
a factor $\sim$0.55 lower and so are the SFRs. Solar metallicity models produce
masses $\sim10\%$ larger. The maximal extinction model with $A_V=0.5$ produces
$\sim45$\% lower ages and $\sim45$\% higher masses (with the consequent increase
in SFR) with more scatter.}
\tablenotetext{a}{Age$_w$ corresponds to the SFH-weighted ages. In the case of
CSF models, this simply corresponds to half the time since the onset of star
formation. The typical uncertainties in this quantity are substantial
($^{+70}_{-120}$ Myr).}
\tablenotetext{b}{The SSFR (= SFR/Mass) here is derived from the extrapolated
$L_{1500}$ luminosity and the masses from the CSF models.}
\tablenotetext{c}{These are not the mean values of the best fit parameters
derived for the sample but rather the parameters derived from the modeling of
the mean SED described in \S6.2 (see figure \ref{meansed}).}
\end{deluxetable*}

Finally, given the challenges in constraining both age and reddening based upon
the photometric information available for individual sources, we will assume
that galaxies in our sample show negligible dust extinction when doing the
stellar population modeling. We have good reasons for making this assumption.
$z\gtrsim5$ galaxies have been found to have very blue $UV$-continuum slopes
$\beta$ and thus little dust extinction (e.g., \citealt{Lehnert2003,
Stanway2005, Yan2005, Bouwens2006b, Bouwens2009a}). As we will show later, a
similarly steep $UV$-continuum slope $\beta$ ($-2.4\pm0.4$) is found for the
mean SED of our sample (\S6.2) -- again suggesting minimal dust extinction. For
individual sources, however, it is difficult for us to obtain useful constraints
on the $UV$-continuum slope and hence dust extinction. While we have high
quality \HH-band fluxes for our candidates, the other fluxes we have which probe
the $UV$-continuum are not adequate: the \JJ-band fluxes we have available
depend significantly on the redshift of the source (the \JJ-band extends to
$\sim$8000~\AA) and the $K$-band fluxes (available on fewer sources) are more
uncertain in general (due to the shallower nature of the $K$-band imaging data).
We have tested the impact of allowing modest amounts of
extinction (A$_V<0.5$) to the models and found that the main results (SMD, SSFR,
ages) are unchanged or at least consistent within the uncertainties.

Within these general specifications, we explored a wide variety of different
parameters (redshift, Age) in modeling the observed photometry of each $z\sim7$
candidate:
\begin{eqnarray*}
z &=& 4.0 - 11.0 {\rm ~(steps~of~0.01)}\\
\log{(Age[Myr])} &=& 7.5 - Age_{max} {\rm ~(steps~of~0.01)} \\
\end{eqnarray*}
where $Age_{max}$ is the age of the universe at the corresponding redshift so
that the models avoid solutions in which the populations are older than the
universe. To derive 68\%, 95\%, and 98\% confidence intervals for the above
parameters, we ran a number of simulations where we added photometric scatter
(noise) to the observed fluxes and then used the results to determine the
threshold $\chi^2$ values that encompass these confidence intervals. We
performed the above calculations with the stellar population modeling code named
FAST (see appendix in \citealt{Kriek2009}).

Note that in stellar population modeling, we do not use the photometric redshift
estimates from the previous section. This was to avoid additional complications
in the definition of the confidence intervals (as the templates are slightly
different). However, we found that the two redshift estimates are consistent
(the typical discrepancies are rms $\Delta z\sim0.04$).

In figure \ref{seds1}, the observed SEDs are presented along with the best fit
models with no extinction. The corresponding properties we derive for these
galaxies using these models are detailed in table \ref{fits}. The ages in that
table correspond to the average age of the stellar population, i.e., $M_{\rm
total}^{-1}\int t_{lookback} SFR(t_{lookback})dt_{lookback}$ (=age$_w$). In the
case of constant SFR models, this is simply equal to one half of the time
elapsed since the onset of star formation. Similar to what others have found at
somewhat lower redshifts, these models indicate the presence of quite massive
systems very early in the universe. We find masses in the range of
$0.2-12\times10^{9}M_{\odot}$. These models also show quite large ages that
place the formation of most of the stars up to 380 Myrs earlier with typical
values in the order of 300 Myr. The uncertainties in the derived age, however,
are quite substantial, typically $^{+70}_{-120}~{\rm Myr}$ (68\% confidence
intervals). From table \ref{fits}, it can be noticed that the values of age$_w$
imply that some galaxies have been forming stars for times comparable to the age
of the Universe at this redshift. We have preferred not to impose any arbitrary
restrictions on the time of onset of star formation because any meaningful
redshift constraint (e.g. z$<$100) would not meaningfully restrict the ages (16
Myr for z $<$100), particularly when compared to the uncertainties associated
with the estimation, typically $^{+70}_{-120}$ Myr. Such constraint implies
insignificant fractions of stellar mass assembled at extreme redshifts.

\subsection{Parameter dependencies} 

In the above stellar population modeling, we adopted a Salpeter IMF and assumed
sub-solar metallicity (0.2Z$_\odot$). We also explored the effects of using a
Chabrier IMF and of varying the metallicity of the models. We find that while
distribution of best fit redshifts and ages is unchanged, the derived masses are
$\sim$45\% smaller if we use Chabrier IMF (instead of Salpeter), and 10\% larger
if we consider models with solar metallicity (instead of 0.2$Z_\odot$). For both
metallicities (0.2 $Z_ {\odot}$ and $Z_{\odot}$) and both IMFs (Chabrier and
Salpeter), we obtain reasonable $\chi^2$ fit results, so there is no reason to
prefer one IMF or metallicity over the others.

We have also assumed that our sources suffer from minimal dust extinction (both
due to the very blue $UV$-continuum slope $\beta$ measured from the mean SED
here [\S6.3] and due to the blue $UV$-continuum slopes $\beta$ observed at
$z\gtrsim5$: \citealt{Lehnert2003, Stanway2005, Yan2005, Bouwens2006b,
Bouwens2009a}). However, this extinction is not very well constrained for
individual $z\sim7$ galaxies (which lack strong constraints on their
$UV$-continuum slopes), and so it is worthwhile to mention how larger values of
the dust extinction would affect the ages and stellar masses derived from our
modeling. Imposing an extinction of $A_V=0.5$ (following
Calzetti et al. 2000 -- we consider this a safe upper limit based on the
previously mentioned studies), yields best fit models that are $\sim40\%$ more
massive (to compensate for the dimming) but $\sim45\%$ younger (which prevents
their $H-[3.6]$ colors from becoming too red to match the observations) with the
consequent increase in SFR. We also find reasonable $\chi^2$ for these models.

Finally, we must remember that we adopted a specific form for the star-formation
history -- supposing the SFR for each was constant in time. We could easily have
adopted other star formation histories (instantaneous burst, exponentially
decaying) in fitting the observed SEDs and found acceptable results. To
determine the approximate effect of the star formation history on our derived
parameters, we also considered exponentially decaying $e^{-t/\tau}$ histories
and instantaneous bursts in modeling our sources. In general we found larger
ages and masses for histories with larger $\tau$'s (where $\tau$ for a constant
SFR model is of course $\infty$ and 0 for an instantaneous burst) though all
assumed histories produced acceptable fits. In that sense, the unrealistic
instantaneous burst models provide a lower limit for the age$_w$ of the sample
that we find to be 80 Myr. Using a more reasonable approximation like $\tau=100$
Myr this number goes up to $170$ Myr. This latest result would imply that we are
observing a quiescent population right after the main star formation episode is
over, which seems unlikely.

While we note these dependencies, we will use our fit results assuming no dust
extinction, a Salpeter IMF, sub-solar metallicity (0.2 $Z_\odot$), and a
constant SFR when deriving results in subsequent sections and in particular to
estimate the Stellar Mass Density (SMD) of the universe.
Relaxing our assumption that there is no dust extinction
(A$_V$=0) to A$_V<0.5$ produces no significant change in the derived
quantities.

\figmeansed

\subsection{Average Spectral Energy Distribution}

One significant challenge in modeling the stellar populations of $z$-dropouts in
our sample is the faintness of the sources and therefore the still sizeable
uncertainties on the fluxes we derive. Consequently, it becomes difficult for us
to obtain tight constraints on the model parameters -- like dust or age -- for
individual sources.

We can obtain much tighter constraints on the properties of our $z\sim7$ galaxy
candidates by averaging the measured fluxes for the sources and deriving a mean
spectral energy distribution, per unit wavelength. This is particularly valuable
for a determination of the $UV$-continuum slope $\beta$ for the sample, since
this slope is constrained from the $H$ and $K_s$-band fluxes and since the
$K$-band flux is only poorly constrained for individual sources.
To derive this mean SED, we first normalize all the sources
to the average \HH-band flux and then take a weighted mean of the fluxes of all
sources ($<x> = \sum (x_i/\sigma_i^2)/(\sum 1/\sigma_i^2)$). The mean $K$-band
flux is derived from the 6 sources where we have deep $K$-band data. The sources
with poor photometry (\gnstwo, \gnsfive\ and \hdfone\ in 3.6\mum) were not
considered when taking the mean. The mean SED is presented in
Table~\ref{photometry} and Figure~\ref{seds1}. The $UV$-continuum slope $\beta$
we estimate from the $H$ and $K$ photometry is $-2.4\pm0.4$.

We perform the stellar population modeling for this source in the same way as
for the other sources and include the estimated properties in Table~\ref{fits}.
When modeling, however, the $z_{850}$ flux has not been
included in the fits to minimize the influence of our lowest redshift sources.
 The best-fit model is also compared with the mean SED in Figure~\ref{seds1}.
If we allow the dust reddening to be non-zero and include that in the fits, we
find a best fit $A_V=0.4$. This provides support for our assumption in \S6 that
our $z\sim7$ galaxy candidates are largely dust free. Using the mass of the best
fit to the mean SED (6.3$\times10^{9}~\rm{M}_\odot$) and the extrapolated
luminosity at 1500 \AA\ (5.5$\times10^{10}~L_\odot$) we infer a mass to light
ratio of $M/L_{UV}$ = 0.12 $M_\odot/L_\odot$, and also a $M/L_{V}$ = 0.28
$M_\odot/L_\odot$ (from L$_{V} = 2.2\times10^ {10}~\rm{M}_\odot$ measured from
the best fit model at 5500 \AA). We will use these M/L ratios to provide one
estimate of the stellar mass density in \S7.

\figSSFRz

\subsection{Specific Star Formation Rate}

A key quantity in considering the build-up of stars within a galaxy is the
specific star formation rate -- similar to the $b$ parameter which was more
frequently used in the past to characterize a galaxy's star-formation history.
It is the star formation rate within a galaxy divided by its stellar mass -- or
equivalently the fraction of the stellar mass in a galaxy that forms per unit
time. As such, the SSFR provides us with a useful way of thinking about galaxy
growth over cosmic time -- and so it is not surprising that it has been
estimated out to $z\sim6$.

Here we make use of our best fit synthetic models to
estimate L$_{1500}$ and then use the usual formula from Madau et al. (1998) to
estimate the SFR. We combine this with the masses of the best fit models to
obtain the specific SFR (see caption in table 3). We find values of 1.8 - 4.1
Gyr$^{-1}$ for this quantity across our sample, with a median value of
$2.4\pm0.6$ Gyr$^{-1}$ (the outlier at $\sim$20 Gyr$^{-1}$ is undetected in the
mid-IR so the constraints are poor). To put this value in context, it makes
sense for us to compare our derived SSFRs with the values at lower redshift.
Given that the SSFRs can depend somewhat on stellar mass, we compare our results
to the same median mass as we find in our sample, $5\times10^9$ $M_{\odot}$.
From the data presented in \citet{Stark2009}, we find 2.1, 2.0, and 2.0
Gyr$^{-1}$ in their $z\sim4,5,$ and 6 samples; from \citet{Papovich2001},
\citet{Sawicki2007} and \citet{Daddi2007} we find $\sim2$ Gyr$^{-1}$ at
$z\sim2$; and from \citet{Noeske2007} we estimate $0.3 -1.2$ Gyr$^{-1}$ for
$z\sim 0.2-1$ samples. At $z\sim2$, \citet{Reddy2006a} find a much higher value
of $\sim10$ Gyr$^{-1}$ but at that mass it is only based on MIPS detected
sources. Obviously, any MIPS-detected sample would be biased to include only
those sources with substantial enough SFR to show MIPS detections, and hence
probably is not representative for the $10^{9.5} M_{\odot}$ population. The
results are shown in Figure \ref{ssfr}.

From $z\sim7$ to $z\sim2$, there is little apparent evolution in the SSFR and
the present results provide a continuation of the trend delineated by
\citet{Stark2007} and \citet{Yan2006} from the \citet{Reddy2006a} points.
However, from $z\sim2$ to $z\sim0$ the SSFR shows a rapid decrease. This
suggests that SFR at $z>2$ mostly proceeds in a largely similar way, but that at
$z<2$ there must be some physical processes inhibiting SFR. Similar to many
other comparisons of merit, e.g., evolution of $M^*$ with redshift
(\citealt{Bouwens2006b, Yoshida2006, Bouwens2007}), we find that SF in galaxies
at $z\gtrsim2$ seems to follow somewhat different principles than for galaxies
at $z\lesssim2$.

\section{Stellar Mass Density}

One of the most fundamental quantities we can try to infer from $z\sim7$
selections and the present stellar population modeling is the stellar mass
density. The stellar mass density tells us how much star formation occurred in
the universe to the point of observation, and therefore provides us with a very
powerful constraint on early galaxy formation.

\subsection{Selection Volumes}

An essential step in determining the stellar mass density from our $z\sim7$
$z$-dropout selection is obtaining an accurate estimate of the selection volume.
This requires that we model the selection of dropout galaxies from all six of
the Bouwens et al.\ (2009, in preparation) search fields and estimate the
effective volume we are able to search versus \HH-band magnitude. The selection
volumes are calculated by adding artificial sources to our search fields and
then attempting to reselect them as $z\sim7$ $z$-dropouts 
according to the criterion described in \S2. The artificial sources are
assumed to have a mean $UV$-continuum slope $\beta$ of $-2$, 
consistent with the observed trends a $4<z<6$ \citep{Bouwens2009a}.
Their pixel-by-pixel surface brightness profiles are identical to those of
random $z\sim4$ $B$-dropouts from the HUDF \citep{Bouwens2007} of similar
brightness but their sizes have been rescaled as $(1+z)^{-1}$, following the
observed size trends with redshift at $z>2$ \citep{Oesch2009d, Ferguson2004,
Bouwens2004a}. We have not included the possible contribution from Lyman alpha
emission to the SEDs. A more detailed description of these simulations can be
found in Bouwens et al.\ (2009, in preparation: but see also
\citealt{Bouwens2008}).

We estimate the following total search volumes as a function of \HH-band
magnitude for the entire Bouwens et al.\ (2009, in preparation ) $z\sim7$
$z$-dropout selection (from both the HST NICMOS and ground-based near-IR data):

\begin{displaymath}
V_{eff} (H_{160}) = \left\{
\begin{array}{ll}
33\times 10^4\,\textrm{Mpc}^3, & \mbox{if $25.3<H_{160}<25.8$} \\ 
16\times 10^4\,\textrm{Mpc}^3, & \mbox{if $25.8<H_{160}<26.3$} \\
6.0\times 10^4\,\textrm{Mpc}^3, & \mbox{if $26.3<H_{160}<27.1$} \\
1.2\times 10^4\,\textrm{Mpc}^3, & \mbox{if $27.1<H_{160}<27.5$} \\
\end{array} \right.
\end{displaymath}

\subsection{Stellar Mass Density Determinations}

We proceed now to estimate the stellar mass density of the sources we find in
these search volumes. We do so in the three different ways we detail below:

\textit{Direct Approach:} Here we estimate the stellar mass density by simply
suming over the expected mass density to come from each source. The expected
mass density is simply the estimated stellar mass for a source multiplied by the
the likelihood it is not a contaminant (90\% for sources in the HUDF and 80\%
otherwise) divided by the selection volume above. The stellar mass density we
derive by summing over the 11 sources is 5.7$\times10^5~M_\odot~{\rm Mpc}^{-3}$.
We estimate the uncertainties in the mass density by bootstrap resampling
(detailed below). This approach has the advantage of being very direct and
even-handedly including all candidates in our selection in the estimate. The
disadvantage, of course, is that this estimate may be affected by the mass
estimates of the three $z$-dropouts with perhaps unreliable 3.6 $\mu$m fluxes.

\textit{Mean SED Approach:} For our second estimate, we use our average SED from
\S6.2 to derive the mean $M/L_{UV}$ ratio for sources in our sample -- reasoning
that this $M/L$ ratio is much more accurately known than any individual $M/L$
ratio in our sample. For our second estimate, we use our average SED from \S6.2
to derive the mean $M/L_{UV}$ ratio for sources in our sample -- reasoning that
this $M/L$ ratio is much more accurately known than any individual $M/L$ ratio
in our sample. If the photometric redshifts are accurate ($z\sim7.2\pm0.5$),
then the limiting depth of our search corresponds to $\sim \rm{M}_{UV,AB}=-20$.
We have then used the \citet{Bouwens2008} $UV$ LF at $z\sim7$ to estimate the
$UV$ Luminosity density integrated to that depth and multiplied it by the
$M/L_{UV}$ ratio to obtain a stellar mass density of 4.5$\times10^5~M_\odot~{\rm
Mpc}^{-3}$. Again for the uncertainties, we rely on a bootstrap resampling
procedure (below).

\textit{Random M/L Approach:} For our final estimate of the stellar mass
density, we ran a monte-carlo simulation where we match up the 11 galaxies in
our $z\sim7$ $z$-dropout sample with randomly sampled $M/L$ ratios from the 8
galaxies with reliable IRAC fluxes. We then divide the masses by the search
volumes that correspond to their UV luminosities. After repeating the match up
process 100000 times we find a median value of 6.6$\times10^{5}~M_\odot~{\rm
Mpc}^{-3}$. The estimate of the uncertainties based on bootstrap resampling is
described below.

\textit{Uncertainty Estimates:} To estimate the uncertainties in our estimate of
the stellar mass density at $z\sim7$, we must fold in the many uncertainties
that contribute to this density, including uncertainties in the sampling of the
LF, the $M/L$ ratios of the galaxies we uncover in our search, whether any
individual source is a contaminant, and finally the mass estimates themselves of
candidates in our sample.

\figSMD

The simplest way of including all these uncertainties in our final estimate is
to run a Monte-Carlo simulation. For each simulation, we iterate over all 11
candidates in our sample and include $P_1(x)$ number of sources in that trial
with a $UV$ luminosity equal to that candidate -- where $P_1 (x)$ is a Poisson
distribution with mean equal to 1. Then, we run over all the sources and give
each source 10-20\% chance of being thrown out (to account for uncertainties in
the contamination fraction). Next, we assign a mass to each of the objects in
the particular realization by drawing a random M/L$_{UV}$ from the observed
values. When doing this, we also include individual uncertainties in the
M/L$_{UV}$ determinations, specifically, we make a weighted choice of a mass
from the distribution associated to a particular source (with the weights
derived from the $\chi^2$ as was already described for the determination of the
ages). Finally, we divide each source by the selection volume appropriate to its
$UV$ luminosity and sum the sources to calculate the stellar mass density for a
given trial. We repeated the simulation 10000 times to ensure that our results
were not limited by the number of trials. After sorting the distribution, we
found that the 68\% upper and lower limits on the stellar mass density were
$3.3\times10^5~M_{\odot}~{\rm Mpc}^{-3}$ and $1.2\times10^6~M_{\odot}~{\rm
Mpc}^{-3}$, respectively.

\textit{Summary:} Above we derive three different estimates of the stellar mass
densities with uncertainties. The estimates are 5.7, 4.5, and
6.6$_{-3.3}^{+5.4}\times10^5~M_\odot~{\rm Mpc}^{-3}$. All these estimates are
consistent with each other but we prefer the random M/L approach because this
one should be less affected by possible systematic errors in the mass derived
from poor photometry. We present this stellar mass density in Figure~\ref{smdz}
and show the previous determinations of \citet{Stark2009} at $z\sim4-6$ and
\citet{Labbe2006} at $z\sim7$ for context. We discuss differences between these
stellar mass density determintions and the observed trends in \S8 and \S9.
We should also note here that allowing moderate extinction
($A_V<0.5$) produces SMD measurements that are fully consistent with the
previous values (within the uncertainties).

\figSFRz

\subsection{SFR Density Determinations}

The advantage of the current stellar population modeling is that it permits to
estimate the SFR in our candidates at even earlier times. Combining our age
constraints with the estimated SMD we can place limits on the SFR density at
even earlier times.

In terms of the previous history, the constant SF models imply ages of
the galaxies consistent with them being in place at around $z\sim10$,
which combined with their assembled masses imply a very simple
estimate of the average SFRD between $7<z<10$. The amount of time
elapsed corresponds to 300 Myr and the stellar mass assembled in that
amount of time is $3.3\times10^5\rm{M}_\odot \rm{Mpc^{-3}}$. This
implies a SFRD of $1.1\times10^{-3}~M_{\odot}~{\rm yr}^{-1}~{\rm
  Mpc}^{-3}$ (see Figure~\ref{sfz}). An extreme approach that would
maximize the SFRD comes from the single burst models. Simply dividing
the total masses ($6.6\times10^5\rm{M}_\odot \rm{Mpc^{-3}}$) by the
ages of the sources (80 Myr for the single burst models) yields an
average SFRD of $8\times10^{-3}~M_{\odot}~{\rm yr}^{-1}~{\rm
  Mpc}^{-3}$ in the previous 80 Myr.

\section{Reliability of Current Results}

\subsection{Comparison with Previous Photometry} 

In general, our optical to near-IR photometry is consistent with the photometry
presented in Bouwens et al. (2009, in preparation) although a systematic offset
of $\sim0.2$ magnitudes is present due to the fact that the fluxes presented
there were measured in a somewhat smaller aperture and no aperture corrections
were applied. We also find excellent agreement between the mid-IR IRAC fluxes
measured for our HUDF $z$-dropouts and those presented in \citet{Labbe2006}.
Although this might not be surprising due to our use of the same technique for
doing photometry, our modeling of the flux from neighboring sources (and
subsequent removal of this flux) is completely independent. This illustrates the
robustness of our method for doing photometry. An independent test of the
quality of photometry can be obtained by comparing flux measurements for sources
with deep IRAC observations taken in both GOODS epochs (rotated by 180 degrees).
In general, we observe excellent agreement between the two measurements for the
five sources with two epoch data (the four UDF $z$-dropouts and \cdfone) --
suggesting that systematics are minimal. The only exception to this is for the
3.6 \mum\ measurement for \gnstwo, \gnsfive, and \hdfone\ where there are bright
nearby neighbors.

\subsection{Comparison with Previous Estimates of the Stellar Mass Density at
$z\sim7$}

Previously, \citet{Labbe2006} made an estimate of the stellar mass density at
$z\sim7$ based upon a small (4 galaxy) HUDF $z$-dropout sample. They estimated a
stellar mass density of 1.6$^{+1.8}_{-0.8}\times10^6~M_\odot~{Mpc}^{-3}$ to
$>$0.3 $L_{z=3}^{*}$. Since we adopt a similar limiting luminosity, we can make
a direct comparison with the stellar mass density estimated here. We find a
fiducial value that is a about half the one estimated by \citet{Labbe2006}. This
is mostly due to our different choice of star formation histories (we favor
constant SF histories versus the average between SSP and constant SFR used in
that work). These estimates are fully consistent within the uncertainties.

This quantity was also estimated by \citet{Stark2009} at $z\sim4-6$ for sources
in the GOODS fields to similar depth (see figure \ref{smdz}). A simple
calculation shows that the observed growth in mass between $z\sim7$ and $z\sim6$
is consistent with the observed SFRD derived from the UV LF studies of
\citet{Bouwens2008}.

\subsection{How Significant Is Crowding for Current Samples}

Because of the broad PSF in IRAC data, crowding is considered to be a
potentially significant concern in doing photometry on faint sources,
particularly when these sources are nearby bright foreground galaxies. In fact,
in many studies, it is thought that perhaps $\sim$50\% of faint sources are
sufficiently close to their neighbors that IRAC photometry is impossible. What
do we find here?

We attempted to do IRAC photometry on all 11 $z$-dropouts in the Bouwens et al.\
(2009, in preparation), without excluding any sources due to crowding issues. Of
the 11 sources presented here, 5 suffered significant blending with nearby
neighbors. However, as a result of our deblending technique \citep{Labbe2006},
we were able to recover reliable fluxes for all 11 in the 4.5 $\mu$m band and 8
of 11 (73\%) in the 3.6 $\mu$m band (excluding \gnstwo, \gnsfive, and \hdfone\
which showed strong residuals from the neighbors after the subtraction process
-- implying large systematic errors in those two cases). This agrees with
Monte-Carlo experiments that we performed that suggest that photometry is
possible for $\gtrsim$80\% of faint sources and that the largest errors in the
recovered fluxes should be roughly a factor of 2.

\section{Discussion}

\subsection{Stellar mass growth during the first Gyr of the Universe.}

Despite the large uncertainties in the derived individual masses, it seems clear
now that quite massive ($>10^{10}~M_\odot$) systems with evolved stellar
populations were already present in the universe at very early times ($z\sim7$
or at least $z\sim6$). These very massive systems likely correspond with the
most massive Dark Matter Haloes (DMH) and are predicted to exist but in low
numbers in the standard Press Schechter formalism. It is hard to run simulations
that can probe this very massive end of the Mass function since to obtain them
in larger numbers it would be necessary to run simulations with quite large
volumes. \citet{Dave2006} ran an SPH simulation with a comoving volume of
$10^5~{\rm Mpc}^{3}$ which they find to be adequate to probe the stellar mass
range $10^{7.2}-10^{10}~M_\odot$. At the median mass of our sample of
$5\times10^9~M_\odot$, they find a number density of $\sim10^{-4}~{\rm
Mpc}^{-3}$. More recently, Choi \& Nagamine (2009, in preparation), ran a
simulation in a larger box of $\sim3\times10^6~{\rm Mpc}^3$ and find the number
density at $5\times10^9~M_\odot$ to be approximately one half that predicted by
\citet{Dave2006}. Given the uncertainties, these predictions agree quite well
with the density of such objects that we find of $1.6 \times 10^{-4}$
Mpc$^{-3}$. Although our sample may still suffer from small number statistics,
the fact is that such massive objects are found to be fairly common at this
early epochs and their moderately evolved SEDs show that their stellar
populations are not pristine but were partly formed at higher redshifts.

\subsection{Star Formation Histories of high redshift galaxies.}

It is well known that the results (e.g., derived masses, SFRs, etc.) of stellar
population analyses can depend significantly on the functional form one assumes
for the star formation history. Some care, therefore, needs to be given to the
parameterization of these histories to ensure that the conclusions drawn do not
depend too much on artifacts of this parameterization.

To illustrate, there are 3 different model parameterizations of the star
formation history in common use in the literature: instantaneous burst models,
exponentially decay star formation models, and constant star formation models.
Instantaneous burst models give younger ages and lower masses than exponential
decay models which in turn give younger ages and lower masses than constant star
formation models. The instantaneous burst models do not seem realistic but are
useful to set lower limits on the derived ages and masses. Exponential decay
models are the most popular in the literature (e.g. \citealt{Yan2006, Eyles2007,
Stark2009}) -- perhaps because of their versatility in modeling a wide range in
star formation histories -- but give rise to a troubling prediction, namely,
that the star formation rates of galaxies are larger in the past. This
prediction is troubling because it contradicts both the observed and predicted
trend that the SFR density of the universe increases with cosmic time.

Given these concerns, we prefer to model the stellar population of high-redshift
galaxies with constant star formation histories. The reason for this preference
is as follows: first, constant star formation models do not naturally predict
that the SFR density will be greater at early times, as exponentially decaying
models do. Second, constant star formation models do not predict a large
population of UV luminous sources at earlier times as found from
exponentially-decaying models (e.g., \citealt{Yan2006}). This is more consistent
with the observation that such sources are not found in large numbers at
$z\sim7$. Third, constant star formation models are more consistent with the low
evolution seen in the specific star formation rate (e.g., Figure \ref{ssfr}
here, \S6.2, and as discussed by \citealt{Stark2009}). We realize that
\textit{some} luminous galaxies may have mass-to-light ratios suggesting their
SFRs were higher in the past, but we suspect this may be a duty cycle issue that
is due to feedback, etc. These galaxies are simply experiencing a period where
their SFRs are less than their norm.

\subsection{Reionization}

As outlined in \S7.3, the stellar population modeling we do of $z\sim7$ galaxy
candidates allow us to estimate the SFR density at even earlier times. Having an
estimate of this SFR density is valuable since it allows us to assess how much
ionizing radiation the star-forming population at $z\gtrsim7$ might likely
produce -- and hence repose the question about whether z$\gtrsim$6 galaxies are
capable of keeping the universe reionized. From absorption studies to bright
$z\sim6$ QSOs, the process of reionizing the hydrogen of the universe was just
ending at $z\sim6$ \citep{Fan2006, Becker2001} while the five year WMAP results
suggest it began at least as early as $z\sim11$ \citep{Komatsu2009}.

\citet{Madau1999} presented a prescription to estimate the critical density of
UV radiation necessary to reionize the universe at a given redshift. Updated to
a more current cosmology \citep{Komatsu2009}, that formula becomes:
\begin{equation}
\small
\rho^{crit}_{SFR}(z)\approx\frac{0.04}{f_{esc}}\left(\frac{1+z}{8}\right)^3
\left(\frac{C} {30}\right) \left(\frac{\Omega_b~h^2_{70}}{0.0463}\right)^2
M_{\odot}~{\rm yr^{-1}~Mpc^{-3}}
\end{equation}
At $z=7$, for an escape fraction $f_{esc}=0.1$ (e.g. \citealt{Shapley2006}), and
for a clumping factor $C=30$ (but see \citealt{Pawlik2009b} who suggest
$C\sim6$) the value of of $\rho^{crit}_{SFR}=0.4 ~M_{\odot}~{\rm
yr^{-1}~Mpc^{-3}}$. The sources capable of producing such radiation, however,
remain unknown but young O and B stars in early galaxies stand out as the most
likely candidates given the observed decrease in the number density of quasars
at high redshifts.

\begin{deluxetable}{cc}
\tiny
\tablecaption{Key results derived from $z\sim$7 sample.\label{results}}
\centering \tablewidth{0pt}
\tablehead{
\colhead{Quantity} & \colhead{Value}
}
\startdata
Redshifts & $7.2\pm0.5$ \\
Masses & 0.1-12$\times10^9$ $M_{\odot}$ \\
$M/L_{UV}$ ratio & 0.01-0.1 $M_\odot/L_\odot$ \\
Minimum Age\tablenotemark{a} & 80 Myr \\
Average Age & 300 Myr \\
$UV$-continuum Slope $\beta$ & $-$2.4$\pm$0.4 \\
Specific SFR & 2.4$\pm$0.6 Gyr$^{-1}$\\
Mass Density (Direct) & 5.7$\times10^5~M_\odot~{\rm Mpc^{-3}}$ \\
Mass Density (M/L of Mean SED) & 4.5$\times10^5~M_\odot~{\rm Mpc^{-3}}$\\
Mass Density (Random M/L)\tablenotemark{b} &
 6.6$_{-3.3}^{+5.4}\times10^5~M_\odot~{\rm Mpc^{-3}}$\\
Predicted SFR density (at $z=9$) &  0.0011 $M_\odot~{\rm yr^{-1} Mpc^{-3}}$ \\
\enddata
\tablenotetext{a}{From single burst models.}
\tablenotetext{b}{Our best estimate.}
\end{deluxetable}

Based upon the stellar mass density and the mean ages we derive in \S6 and \S7
for the sample, we estimated an average SFR density of 0.0011 $M_\odot~{\rm
yr^{-1}~Mpc^{-3}}$ between $7<z<10$ (\S7.3). This is more than 2 orders of
magnitude below the SFR density required to reionize the universe at $z=7$. We
can also obtain an upper limit to the SFR density by considering the minimal
ages obtained from the single burst models. The mean age of 80 Myr obtained from
these models implies a SFR density of 0.008 $M_\odot~{\rm yr^{-1}~Mpc^{-3}}$ at
$z=8$ (\S7.3). Even for this larger value for the SFR density, we are still a
factor $\sim$50 below that required to reionize the universe. Of course, these
values are based only on the brightest observable sources and so including the
mass from galaxies of even lower luminosities than our selection limit would
increase these numbers by a factor of $\gtrsim2-3$. In any case, it bears
mention that we cannot include the contribution from a population of dust
obscured sources that we would miss in LBG selections. However, as we have
already noted in \S6.1 and \S6.2 there is strong evidence that the contribution
from this population is not large (e.g., \citealt{Bouwens2009a}).

\section{Summary}

We use the very deep optical, near-IR, and IRAC data over and around the two
GOODS fields to study the properties of a large sample of $\sim$11 $z$-dropout
galaxies at $z\sim7$. Considered are the ages, stellar masses, redshifts, and
dust properties of $z\sim7$ galaxies. The $z$-dropout candidates were drawn from
the very large selection of such galaxies from $\sim$80 arcmin$^2$ of deep
NICMOS data by Bouwens et al.\ (2009, in preparation). Essential to this
analysis is the availability of deep Spitzer IRAC mid-IR data that give us deep
coverage of these sources at rest-frame optical wavelengths and hence permits us
to estimate the stellar mass. The stellar population modeling was performed with
the \citet{Bruzual2003} spectral synthesis modeling code with a Salpeter IMF and
assuming solar metallicity. We have quantified how our best-fit properties would
change, as a function of the assumed IMF and metallicity.

Our conclusions are as follows:

\begin{itemize}

\item Photometric redshifts place the candidates at $6.2<z<8.0$ with the mean
redshift of the sample at 7.2. The uncertainties in the redshift for individual
candidates are typically $\Delta z$ $\sim$ 0.5 and have been taken into account
in the derivation of the confidence intervals for the rest of the properties
(see \S5 and figure \ref{zdist}).

\item We use the results of of \citet{Bouwens2009a} regarding the low
extinctions expected at high redshifts to help us better constrain the
individual ages of the sources. Our CSF model fits yield SFRs that are
consistent with simple conversion based on the L$_{1500}$ luminosities. The
best-fit ages allow enough time to assemble their masses by the redshift of
observation. This does not seem to be the case for the most massive sources in
the work by \citet{Yan2006}. We argue that this is not likely an effect of large
extinctions but rather that these sources are the result of some other more
complicated history probably involving mergers (\S9.2).

\item The star formation history weighted ages of the observed stellar
population we derive are in the range of $170-420$ Myr with a mean of 300 Myr
(there is an outlier with $\sim20$ Myr). These ages are consistent with previous
work that place the formation of the most massive galaxies at very early epochs.
In particular, the bulk of the stars in some of these sources seem to have
formed as early as $z\sim10$ (see \S6 and table \ref{fits}).

\item The stellar masses we estimate for individual $z\sim7$ $z$-dropouts in our
sample range from 0.2$\times10^9$ $M_\odot$ to 12 $\times10^9$ $M_\odot$, with a
mean for the sample of 5.1$\times10^9$ $M_\odot$. The masses we estimate are
much more well constrained than other quantities -- like the age -- but are
nevertheless still uncertain at the factor of 2 level (\S6 and table 
\ref{fits}).
  
\item We find that the specific SFRs (SFR/Mass) of the sources in the sample
range from 1.8 Gyr$^{-1}$ to 25 Gyr$^{-1}$, with a biweight mean value of
2.4$\pm$0.6 Gyr$^{-1}$. We observe that at comparable masses, the specific SFR
is surprisingly close to all the values in the literature between $z\sim2$ and
$z\sim6$ (\citealt{Reddy2006a, Papovich2001, Daddi2007, Stark2009}). The
constancy of this quantity between $z\sim2-7$, in contrast with its fast decline
at $z\lesssim2$ (\citealt{Noeske2007}), suggests that star formation proceeds in
different ways in these two regimes (see \S6.3 and figure \ref{ssfr}).

\item Utilizing the estimated selection volumes for the Bouwens et al.\ (2009,
in prep) $z$-dropout search, we derive the stellar mass density at $z=7$ using
an approach that randomly samples the M/L of the galaxies with the most reliable
IRAC photometry. We find 6.6$_{-3.3}^{+5.4}\times10^5~M_\odot~{\rm Mpc^{-3}}$.
The random M/L approach is preferred as it is less affected by possible
systematic errors in photometry. We tested other approaches, including averaging
the direct fits, yielding very similar answers. Our estimate of the global
stellar mass assembled is consistent with the growth expected based on the SFRD
measured between $6<z<7$ and the SMD measured to similar depths at $z\sim6$
(\S7.2).

\item Combining the estimated ages and total assembled mass we can derive an
average SFR between $z\sim10$ and $z\sim7$ of $1.1\times10^{-3}~M_{\odot}~{\rm
yr^{-1}~Mpc^{-3}}$. SSP models provide us with minimum age estimates that in
combination with the masses allow us to place an upper limit to the SFRD at
$z\sim8$. This estimate is still a factor 50 below the necessary value to
reionize the universe at this redshift (following the \citealt{Madau1999}
prescription with $f_{esc}=0.1$ and $C=30$, eq. 1), in agreement with previous
works. We emphasize that this estimate is only based on the most luminous
sources and is probably missing most of the UV light, which is being produced by
sources in the faint end of the LF (e.g., \citealt{Bouwens2007, Reddy2009,
Yan2004a}).
\end{itemize}

The Bouwens et al. (2009, in prep) search based on a large area of high quality
optical to NIR data has provided us with the first sizable sample of candidate
sources at $z\sim$7, a step of 200 Myr with respect to the previous efforts (but
see also new work by \citealt{Oesch2009a, Bouwens2009b, McLure2009b,
Bunker2009, Ouchi2009}). Complementing these data with very deep mid-IR
Spitzer IRAC imaging we have been able to fit BC03 SSP models to estimate the
masses and ages of galaxies at $z\sim7$. Our results suggest that these galaxies
had been forming stars for $\gtrsim$200 Myr and as soon as $z$$\sim$9-10, well
into the reionization epoch. We expect to substantially improve upon these
results taking advantage of the deep near-IR data soon to become available over
the HUDF and CDF-South GOODS field as a result of the new WFC3 instrument on
HST. Not only will we substantially increase the number of $z\sim7-8$ galaxies
known, but the deeper data and improved set of near-IR filters $(Y, J, H)$ will
enable us to perform much more accurate stellar population modeling on
individual sources.

\begin{deluxetable*}{lccccccccc}
\tabletypesize{\footnotesize}
\tablecaption{Summary of Photometry. \label{photometry2}}
\centering
\tablewidth{0pt}
\tablehead{
\colhead{ID}&\colhead{\BB}&\colhead{\vv}&\colhead{\ii}&\colhead{\zz}&
\colhead{\JJ}&\colhead{\HH}&\colhead{$K_s$}&\colhead{3.6\mum}&
\colhead{4.5\mum}\\
}
\startdata
\udfone&0.18$\pm$0.41&-0.22$\pm$0.26&-0.07$\pm$0.41&2.88$\pm$0.63&8.91$\pm
$1.37&
13.13$\pm$2.11&17.49$\pm$3.14&30.88$\pm$3.51&16.75$\pm$6.88\\
\udftwo&-0.37$\pm$0.41&-0.07$\pm$0.26&-0.30$\pm$0.44&0.04$\pm$0.74&7.77$\pm
$1.37&
7.06$\pm$1.66&8.25$\pm$4.66&12.06$\pm$3.51&8.21$\pm$6.88\\
\udfthree&-0.07$\pm$0.37&0.04$\pm$0.22&-0.04$\pm$0.37&0.85$\pm$0.63&3.11$\pm
$1.04&
6.32$\pm$1.63&2.03$\pm$4.07&2.96$\pm$3.55&0.07$\pm$6.99\\
\udffour&-0.41$\pm$0.37&-0.22$\pm$0.26&-0.74$\pm$0.37&-0.59$\pm$0.67&3.48$\pm
$1.07&
6.32$\pm$1.78&-1.26$\pm$5.51&9.47$\pm$3.55&7.40$\pm$6.92\\
\gnsone&-0.78$\pm$1.33&-0.74$\pm$0.89&-0.30$\pm$1.78&1.33$\pm$2.48&9.32$\pm
$2.29&
12.50$\pm$2.55&\nodata&40.20$\pm$5.07&26.11$\pm$9.69\\
\gnstwo&-0.41$\pm$1.04&0.89$\pm$0.81&-2.29$\pm$1.59&1.29$\pm$1.59&5.66$\pm
$1.96&
7.73$\pm$2.48&\nodata&9.84$\pm$5.33&11.87$\pm$9.80\\
\gnsthree&-0.55$\pm$0.85&0.33$\pm$0.67&-0.33$\pm$0.96&0.44$\pm$1.26&4.25$\pm
$2.66&
7.73$\pm$2.33&\nodata&18.97$\pm$5.07&15.87$\pm$9.65\\
\gnsfour&-0.15$\pm$1.48&-0.11$\pm$1.07&-1.04$\pm$1.59&1.41$\pm$1.63&7.58$\pm
$2.48&
11.13$\pm$3.00&\nodata&31.36$\pm$4.77&26.70$\pm$10.21\\
\gnsfive&-0.33$\pm$1.07&0.11$\pm$0.81&-1.89$\pm$1.22&2.15$\pm$1.41&11.76$\pm
$3.37&
28.18$\pm$3.48&24.59$\pm$4.07&52.74$\pm$3.66&25.78$\pm$7.06\\
\cdfone&0.78$\pm$1.26&0.59$\pm$0.92&-2.77$\pm$1.33&1.59$\pm$1.63&6.77$\pm
$2.37&
9.76$\pm$2.70&\nodata&13.24$\pm$3.55&14.68$\pm$7.06\\
\hdfone&-1.26$\pm$0.74&0.59$\pm$0.67&1.11$\pm$1.33&9.21$\pm$1.89&12.39$\pm
$2.55&
15.94$\pm$2.22&10.02$\pm$3.40&30.18$\pm$6.99&19.45$\pm$11.80\\
\enddata
\tablecomments{Photometry for the sample.  Fluxes are in units of 10 nJy.  This 
table is equivalent to table \ref{photometry} but here we present the actual 
flux measurements in the optical instead of tabulating the $1\sigma$ upper 
limits.}
\end{deluxetable*}

\acknowledgments
We would like to thank the referee for his/her very detailed
and insightful report. It has helped us to greatly improve the presentation of
our work. We would also like to show our appreciation to Ken Nagamine and
Junhwnan Choi for giving us early access to the latest results from his
simulations. We also acknowledge useful discussions with Daniel Stark. V.G. is
gratefully for the support from a Fulbright-CONICYT scholarship. This work is
based, in part, on observations made with the Spitzer Space Telescope, which is
operated by the Jet Propulsion Laboratory, California Institute of Technology
under a contract with NASA. We acknowledge the support of NASA grant NAG5-7697.

\appendix
Table \ref{photometry}, in the main text, summarizes the photometry for the
sample in the usual AB magnitude system. Since this is a z-dropout sample, most
of the optical measurements correspond to upper limits which makes it hard to
reproduce the stellar population modeling results presented in this work. In
table \ref{photometry2} of this appendix we provide an equivalent table with the
actual flux measurements in physical units.

\end{document}